\documentclass[prl,twocolumn]{revtex4}
\usepackage{dcolumn}
\usepackage{multirow}
\usepackage{graphicx}
\usepackage{amssymb}
\usepackage{bm}
\usepackage{hyperref}
\usepackage{epstopdf}
\usepackage{color}
\usepackage{mathrsfs}
\usepackage{amsmath,amssymb,amsthm}
\usepackage{rotating}
\usepackage{sverb, longtable}
\usepackage{subfigure}

\usepackage[graphicx]{realboxes}
\usepackage{adjustbox}

\newcommand{\beginsupplement}{%
        \setcounter{section}{0}
        \renewcommand{\thesection}{S\arabic{section}}%
        \setcounter{table}{0}
        \renewcommand{\thetable}{S\arabic{table}}%
        \setcounter{figure}{0}
        \renewcommand{\thefigure}{S\arabic{figure}}%
     }

\begin{document}

\title{Lighting Dark Ages with Tomographic ISW Effect}
\author{Deng Wang}
\email{dengwang@ific.uv.es}
\affiliation{Instituto de F\'{i}sica Corpuscular (CSIC-Universitat de Val\`{e}ncia), E-46980 Paterna, Spain}
\author{Olga Mena}
\email{omena@ific.uv.es}
\affiliation{Instituto de F\'{i}sica Corpuscular (CSIC-Universitat de Val\`{e}ncia), E-46980 Paterna, Spain}
\begin{abstract}
The integrated Sachs-Wolfe effect (ISW) describes
how CMB photons pick up a net blue or redshift when traversing the time-varying gravitational potentials between the last scattering surface and us. Deviations from its standard amplitude could hint new physics. We show that reconstructing the amplitude of the ISW effect as a function of the redshift may provide a unique tool to probe the gravity sector during the era of dark ages, inaccessible via other cosmological observables. Exploiting Planck CMB temperature, polarization and lensing observations, we find a $2\sigma$ deviation from the standard ISW amplitude at redshift $z=500$. Barrying a systematic origin, our findings could point to either possibly new physics or a departure from the standard picture of structure formation under the General Relativity framework. Assuming the simplest two-redshift-bin scenario, we ensure $38\sigma$ and $2\sigma$ evidences of the early and late ISW effects, respectively, despite a priori possible degeneracy with the CMB lensing amplitude. Using a multiple tomographic method, we present the first complete characterization of the ISW effect over space and time.  
Future tomographic ISW analyses are therefore crucial to probe the dark ages at redshifts otherwise unreachable via other probes.

\end{abstract}
\maketitle

\textit{Introduction.} The cosmic dark ages, which encode a wealth of cosmological information, lie between the recombination epoch when the CMB photons decoupled from matter at $z\sim1090$ and the appearance of the first light sources at $z\sim30$. The line absorption due to the 21cm spin-flip transition in neutral hydrogen can extract physical information at $30\lesssim z \lesssim 300$ \cite{Lewis:2007kz}. However, an important question is how to dig up useful information at $300\lesssim z \lesssim 1090$ in current cosmological observations. A viable solution is the integrated Sachs-Wolfe (ISW) effect ~\cite{Sachs:1967er}, one class of secondary CMB temperature fluctuations, which is a powerful probe of dark energy, the nature of gravity, neutrinos and other extra degrees of freedom present at recombination. This effect is originated from the interaction between CMB photons and the time-dependent gravitational potentials along the line of sight between us and the last scattering surface. Given the fact that the gravitational potentials will remain constant unless the expansion of the universe is not entirely driven by a non-relativistic matter component, there can be identified two ISW effects at two distinct periods in the expansion of the universe. Namely, the early ISW effect is sensitive to relativistic particles at the recombination epoch, such as neutrinos, axions or other cosmic relics contributing to the dark radiation of the universe. On the contrary, the late ISW effect is a
test of the late-time density perturbations, depending on the dark energy evolution, on modifications of the gravity sector, or on any other modification to the growth of perturbations (curvature, massive neutrinos, among others). 
Notice that since the ISW effect is always subdominant with respect to primary sources in the CMB, it has to be measured by cross correlating the large scale structure (LSS)~\cite{Crittenden:1995ak,Zaldarriaga:1998te}, or from CMB polarization induced by intra-cluster electron scattering~\cite{Cooray:2003hd}.  Progresses in precise CMB measurements and large scale galaxy surveys enable
detections of the ISW-LSS cross correlation and confirm the decay of the cosmological gravitational potential~\cite{Fosalba:2003ge,Afshordi:2003xu,SDSS:2003lnz,Boughn:2003yz,Fosalba:2003iy,WMAP:2003gmp,Padmanabhan:2004fy,Giannantonio:2006du,Rassat:2006kq,Ho:2008bz,Giannantonio:2008zi,Manzotti:2014kta,Creque-Sarbinowski:2016wue,Stolzner:2017ged,Bahr-Kalus:2022yny,Shajib:2016bes}. The Planck collaboration~\cite{Planck:2015fcm} reports a $4\sigma$ detection by cross-correlating CMB temperature maps with different LSS tracers. Reference~\cite{Cabass:2015xfa} shows a $36\sigma$ detection of the early ISW effect using Planck 2015 CMB data and finds a correlation between the early ISW amplitude and the lensing amplitude, interfering with the measurement of the ISW amplitude. 21cm surveys \cite{Ahn:2023spf} can also be used to detect the ISW effect, for instance, cross correlating their measurements at high redshifts with galaxies \cite{Raccanelli:2015lca}. Moreover, ISW effects are studied in a number of dark energy and alternative gravity models \cite{Cai:2013toa,Smith:2023fob,Seraille:2024beb}.

\begin{figure}
	\centering
	\includegraphics[scale=0.5]{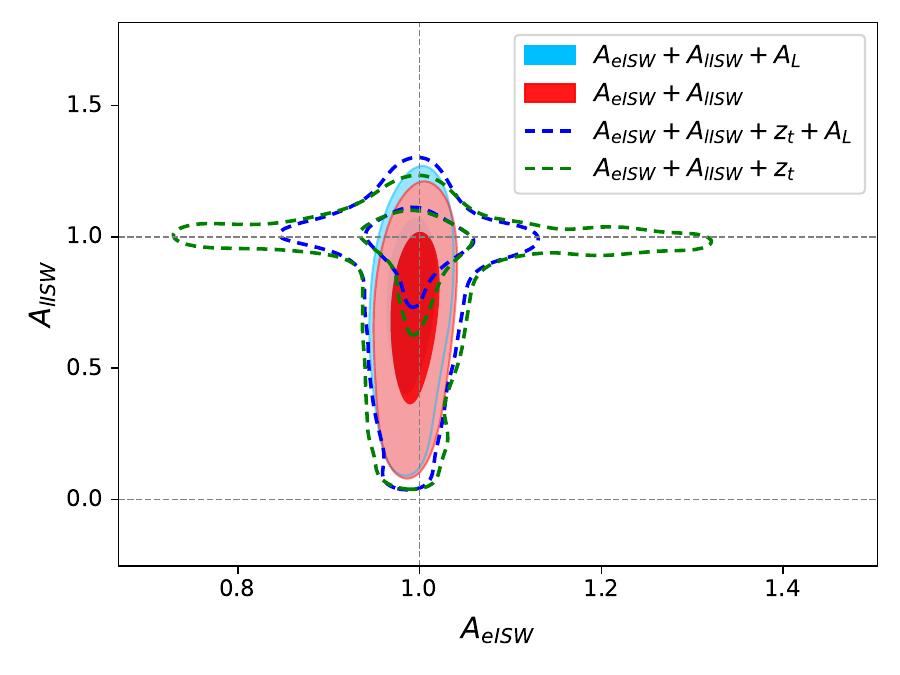}
	\caption{$68\%$ and $95\%$~CL allowed contours in the ($A_{\rm{eISW}}$,\,$A_{\rm{lISW}}$) plane from Planck CMB data with $A_{\rm L}=1$ or free $A_{\rm L}$ in both $z_t=30$ (filled) and free $z_t$ (unfilled) ISW models.}
 \label{fig:f1}
\end{figure}

In this \emph{letter}, we shall consider several tomographic parameterizations of the ISW effect, finding that  CMB data only are able to measure the late ISW effect with a significance above $2\sigma$ independently of parameter degeneracies, as shown in Fig.~\ref{fig:f1}. The novelty of our study also resides on the exploration of the redshift and scale dependence of the ISW effect. Crucially, we demonstrate that our tomographic method, reconstructing the ISW effect amplitude as a function of redshift, can serve as a powerful probe of the dark ages, inaccessible via other observables. A larger value of $A_{\rm ISW}>1$ points to a stronger time variation of the gravitational potentials during the dark ages. Indeed, we find a $2\sigma$ deviation of the ISW amplitude ($A_{\rm ISW} >1$) at redshift $z=500$,  which could be signaling either new physics beyond the $\Lambda$CDM cosmology, a signal beyond the standard structure formation, or unidentified systematical errors.

\textit{Basics.}
The ISW effect leads to a CMB temperature perturbation given by the line-of-sight integral between us and the last scattering surface:
\begin{equation}
\label{eq:isw}
\Theta_\ell (k,\eta_0) = \int_0^{\eta_0} d\eta e^{-\tau(\eta)}[\dot{\Psi} (k, \eta) -\dot{\Phi} (k, \eta)] j_\ell [k(\eta_0-\eta)]~,
\end{equation}
\noindent 
where dot refers to conformal time derivative, $\tau$ is the optical depth, $\eta_0$ is the present  conformal time and $\Psi$ and $\Phi$ are the gravitational potentials in the Newtonian gauge. Before recombination the factor $e^{-\tau(\eta)}$ is negligible, vanishing therefore the ISW effect.

\textit{Data and methodology.} 
CMB observations can measure accurately the matter components, the gravity sector, the topology and the large scale structure of the universe.
To probe the high redshift ISW effect, we mainly use the Planck 2018 high-$\ell$ \texttt{plik} temperature (TT) likelihood at multipoles  $30\leqslant\ell\leqslant2508$, polarization (EE) and their cross-correlation (TE) data at $30\leqslant\ell\leqslant1996$, and the low-$\ell$ TT \texttt{Commander} and \texttt{SimAll} EE likelihoods at $2\leqslant\ell\leqslant29$ \cite{Planck:2019nip,Planck:2018vyg}. We employ conservatively the Planck lensing  likelihood~\cite{Planck:2018lbu} from \texttt{SMICA} maps at $8\leqslant\ell \leqslant400$. As a comparison, we also consider the Atacama Cosmology Telescope (ACT) DR4 TTTEEE data \cite{ACT:2020frw,ACT:2020gnv} at $350<\ell<8000$ and the South Pole Telescope (SPT-3G) TTTEEE likelihood \cite{SPT-3G:2021eoc,SPT-3G:2022hvq} at $750\leqslant\ell<3000$. 

To study the high-$z$ cosmological physics, we divide the full ISW effect into different redshift bins in Eq.~(\ref{eq:isw}) and set the ISW amplitude in each bin as a free parameter. We use these amplitudes to characterize the strengths of the ISW effect and the possible deviations from $\Lambda$CDM through the cosmic history. Note that the theoretical prediction of the amplitude in each bin is 1 under $\Lambda$CDM. Specifically, in order to depict the ISW effect, we study ten scenarios: (i) a global amplitude $A_{\rm{ISW}}$; (ii) early and late ISW effects depicted by two amplitude parameters,  $A_{\rm{eISW}}$ and $A_{\rm{lISW}}$, and a transition redshift $z_t=30$; (iii) same as (ii) but setting the CMB lensing amplitude $A_{\rm L}$ free; (iv) same as (i) but setting $z_t$ free; (v) large and small scale ISW effects described by $A_{\rm{lkISW}}$, $A_{\rm{skISW}}$ and a transition scale $k_t$; (vi) a 5-parameter model containing $z_t$, $k_t$, $A_{\rm{eISW}}$, $A_{\rm{lkISW}}$ and $A_{\rm{skISW}}$, where we divide the large scale ISW effect described by $A_{\rm{lISW}}$ into two parts denoted by $A_{\rm{lkISW}}$ and $A_{\rm{skISW}}$;
(vii) 12 redshift bins located at the $z$ nodes in the range $z\in[0,\,1600]$: $z=10, 30, 50, 80, 150, 300, 500, 700, 900, 1100, 1300$ and $1500$; (viii) an oscillating ISW amplitude $A_{\rm ISW}(z)=A_{base}+A_{amp}[\sin(\omega_{isw}z)+\phi_0]$, where $A_{base},\,A_{amp},\,w_{isw}$ and $\phi_0$ depict the basic ISW amplitude, oscillating strength, frequency and initial phase of the oscillation, respectively; (ix) 32 redshift bins in the same range as (vii); (x) 8 bins in $k$ space, see the Supplementary Material (SM).  

We implement our tomographic ISW models in the publicly available Boltzmann solver \texttt{CAMB} \citep{Lewis:1999bs} and employ the Monte Carlo Markov Chain (MCMC) method to infer the posterior distributions of model parameters by using the package \texttt{CosmoMC}~\cite{Lewis:2002ah,Lewis:2013hha}. We analyze the MCMC chains via the public package \texttt{Getdist} \cite{Lewis:2019x}. The convergence diagnostic of the MCMC chains is the Gelman-Rubin quantity $R-1\lesssim 0.02$ \cite{Gelman:1992zz}.  We take the following uniform priors for model parameters: the baryon fraction $\Omega_bh^2 \in [0.005, 0.1]$, the cold dark matter fraction CDM fraction $\Omega_ch^2 \in [0.001, 0.99]$, the acoustic angular scale at recombination $100\theta_{MC} \in [0.5, 10]$, the amplitude of primordial power spectrum $\mathrm{ln}(10^{10}A_s) \in [2, 4]$, the scalar spectral index $n_s \in [0.8, 1.2]$, the optical depth $\tau \in [0.01, 0.8]$, $A_{\rm L} \in [0, 2.5]$, $z_t \in [0, 1100]$, $k_t \in [10^{-8}, 10^{2}]$ $h$ Mpc$^{-1}$, $A_{amp} \in [-2, 2]$, $\omega_{isw} \in [-10^3, 10^3]$, $\phi_0 \in [-2, 2]$
and all the ISW amplitudes are varied in the range $[-5, 5]$.

\textit{Results.} Figure ~\ref{fig:f1} shows the constraints in the ($A_{\rm{eISW}}$,\,$A_{\rm{lISW}}$) plane from Planck CMB data \cite{Planck:2018vyg,Planck:2019nip,Planck:2018lbu} with a lensing amplitude $A_{\rm L}=1$ or free lensing amplitude $A_{\rm L}$ in both the $z_t=30$ and the free $z_t$ ISW models.
For the case of $z_t=30$, we identify a $38\sigma $ and a $2\sigma$ extraction of the early and late ISW effects, respectively. Due to focusing on relatively small scales, ACT only gives $6\sigma$ and $2\sigma$ measurements with larger errors than Planck, and SPT-3G can not give any constraint (see the SM). If we instead consider one global amplitude for the ISW effect, we find a $37\sigma$ detection from Planck observations, while ACT DR4 ~\cite{ACT:2020frw,ACT:2020gnv} and SPT-3G~\cite{SPT-3G:2021eoc,SPT-3G:2022hvq} give $\sim 6\sigma$ and $4\sigma$ detections of the ISW effect, respectively. The mean values with $1\sigma$ uncertainties from Planck data analyses are $A_{\rm{eISW}} = 0.995\pm 0.026$ and $A_{\rm{lISW}} = 0.67^{+0.35}_{-0.28}$. Splitting the ISW effect into an early and a late-time contribution does not imply an enhancement of the error on its amplitude, if parameterized as a single amplitude $A_{\rm ISW}$. Figure~\ref{fig:f4} in the SM depicts the one-dimensional probability distributions and the two-dimensional contours for the most interesting cosmological parameters here parameterizing the ISW effect via one single amplitude $A_{\rm ISW}$. Notice that the error is almost identical to the one obtained in $A_{\rm{eISW}}$ in the two-parameter model. Indeed, we find $A_{\rm ISW} = 0.997\pm 0.027$ at $68\%$ confidence level (CL). 

\begin{figure*}[htbp]
\centering
\includegraphics[scale=0.45]
{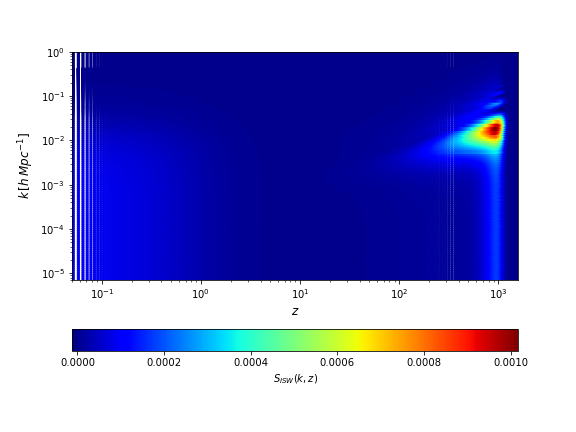} 
\includegraphics[scale=0.35]
{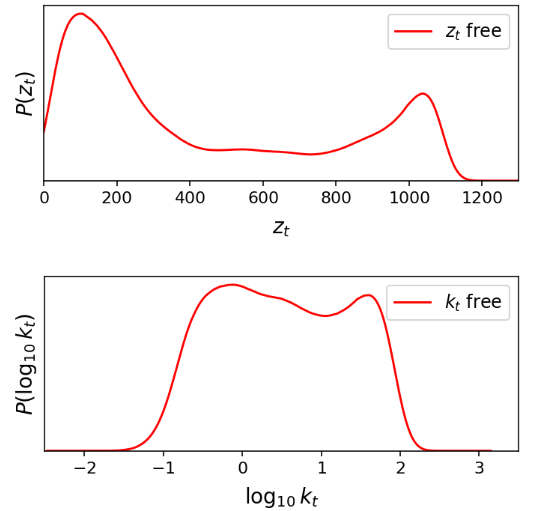}
 	\caption{{\it Left.} The ISW source $S_{\rm{ISW}}(k,\,z)$ as a function of redshift $z$ and scale $k$. {\it Right.} One-dimensional posterior distributions of the transition redshift $z_t$ ({\it top}) in the $A_{\rm{eISW}}+A_{\rm{lISW}}+z_t$ case and of 
  the transition scale $k_t$ ({\it bottom}) in the $A_{\rm{lkISW}}+A_{\rm{skISW}}+k_t$ case from Planck CMB data.}
   \label{fig:f2}
\end{figure*}

More importantly, the ISW detection and its significance do not depend on the lensing amplitude $A_{\rm L}$ (see the overlapping red and blue contours in Fig.~\ref{fig:f1}), despite the fact that the CMB lensing potential is highly correlated with the ISW signal because both effects reside on the very same gravitational potential. We find $A_{\rm{eISW}} = 0.991\pm 0.026$, $A_{\rm{lISW}} = 0.71^{+0.37}_{-0.28}$, that is, the early (late) ISW effect is detected with a significance of $38\sigma$ ($2\sigma$). Very interestingly, the lensing amplitude is perfectly consistent with $A_{\rm L}=1$,  $A_{\rm L} = 1.065^{+0.046}_{-0.060}$, alleviating the so-called  lensing anomaly~\cite{Planck:2018vyg,Motloch:2018pjy}.  In the absence of the ISW rescaling here introduced, Planck CMB data shows a \emph{preference} for additional lensing, leading to $A_\mathrm{L} > 1$ at 3$\sigma$ from temperature and polarization anisotropies. Adding lensing information somewhat diminishes the tension, albeit the value of the lensing amplitude is still above the canonical one by about $2\sigma$.

Two straightforward questions would be what $z_t$ value is preferred by data and how $z_t$ affects the measurement of ISW effect, concerning the 2-bin model here exploited. We obtain the $1\sigma$ constraint $z_{t} = 434^{+700}_{-400}$ from Planck data, which is obviously inconsistent with $z_t=30$. This indicates that the transition redshift is larger than previously expected. The right upper panel of Fig.~\ref{fig:f2} shows the one-dimensional posterior distribution for $z_t$, which is bimodal, with two distinct peaks at high and low redshifts, clearly pointing to the period where the early and late ISW efects are expected to be located, one at $z\simeq 1000$ and the other one at $z\simeq100$, respectively. The detection of both the early and late ISW effects studied here is therefore strongly robust and solid, and independent of the precise parameterization details, and highly independent also of other related rescaling parameters, such as the lensing amplitude $A_{\rm L}$.  In Fig.~\ref{fig:f1}, for $A_{\rm L}=1$ case, the $1\sigma$ constraining values are $A_{\rm eISW} = 1.003^{+0.034}_{-0.050}$ and $A_{\rm lISW} = 0.83^{+0.26}_{-0.11}$ respectively, ensuring a $20\sigma$ and a $3\sigma$ detection of the ISW effect. 
For free $A_{\rm L}$ case, the corresponding values reconstructed are: $A_{\rm eISW} = 0.994\pm 0.045$,  $A_{\rm lISW} = 0.84^{+0.28}_{-0.14}$ and $A_{\rm L} = 1.066^{+0.050}_{-0.059}$, implying a $22\sigma$ and a $3\sigma$ statistical detection of the early and late ISW effects, respectively, and rendering the lensing amplitude anomaly to a simple statistical fluctuation.

Given the fact that the ISW amplitude clearly shows a non-trivial redshift dependence, it is mandatory to test whether $A_{\rm ISW}$ has a non-negligible scale dependence. For that purpose, we assume the possibility of a transition scale $k_t$ in the $k-$space (fixing $z_t=30$). In the left and lower right panels of Fig.~\ref{fig:f2}, we present, respectively, the CMB ISW source function $S_{\rm{ISW}}(k,\,z)=2\dot{\Phi}(k,\,z)e^{-\tau(z)}$ assuming the Planck 2018 best fit cosmology and vanishing anisotropic stress, and the one-dimensional posterior distribution of ${\rm log}_{10}\,k_t$. The theory predicts that the largest ISW effect occurs around $k\sim10^{-2}h$~Mpc$^{-1}$ , corresponding to cluster and supercluster scales, and appears around $z\sim1000$. We obtain a good $1\sigma$ constraint ${\rm log}_{10}\,k_t = 0.52\pm 0.85$ in $k$ range, implying that the ISW effect mainly works on large scales ${\rm log}_{10}\ (k_t < -0.33$ at $1\sigma$ CL). This is very consistent with the theoretical prediction. The constraining values $A_{\rm lkISW} = 0.998\pm 0.027$ and $A_{\rm skISW} = -0.1^{+2.5}_{-3.1}$ show a $37\sigma$ detection of the large scale ISW effect and zero contribution from small scales, respectively. 
One can easily see that we implement a complete characterization of the ISW effect over time and space in Fig.~\ref{fig:f2}.  
Moreover, a joint analysis including both $k_t$ and $z_t$ as free parameters is not a simple superposition from the $z_t$-only and the $k_t$-only analysis. It gives a similar constraint on $k_t$ but a single peak of $z_t$, which is not unexpected and subject to our multiple tomographic modelling, 
see also Fig.~\ref{fig:f8} in the SM, where we present the ISW amplitudes for eight different nodes in the $k$ logarithmic scale, see Fig.~\ref{fig:f10}. That is, we have performed bins in logarithmic space in $k$, in order to reassess the scale dependence of the ISW effect. Despite we clearly notice a turn-on of the ISW effect at scales around $10^{-4}h$~Mpc$^{-1}$ and a turn-off at scales around $1h$~Mpc$^{-1}$, all the ISW amplitudes for the different $k$-nodes are perfectly compatible with the standard value $A_{\rm ISW}=1$. Therefore, our tomographic scale analysis demonstrates that the ISW effect is perfectly compatible with the $\Lambda$CDM prediction on large scales. This is also consistent with what we have learned in the free $k_t$ case.

\begin{figure}
	\centering
	\includegraphics[scale=0.6]{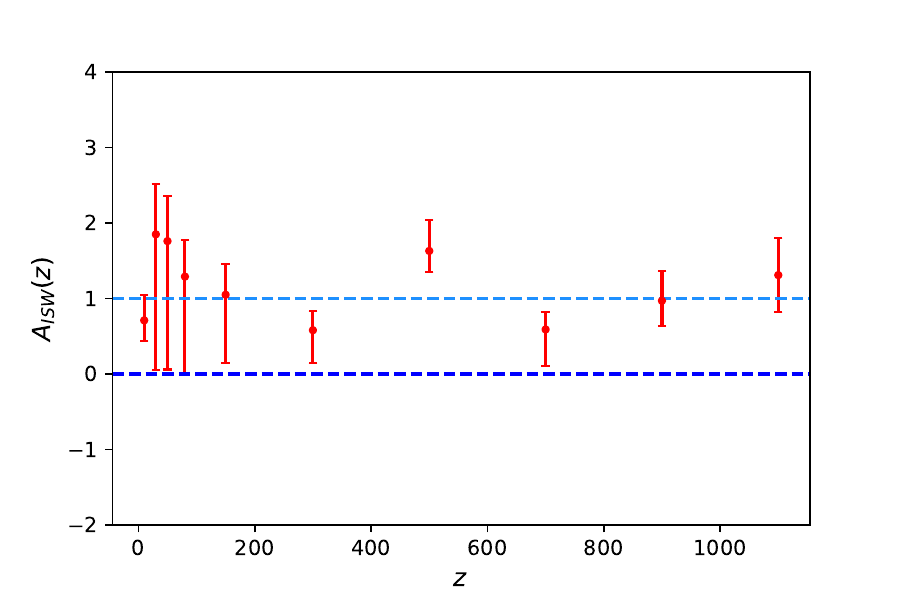}
	\caption{The tomographic redshift analysis of the ISW effect amplitude in the 12-bin model. The data points with error bars denote the mean values of the ISW amplitudes with $1\sigma$ uncertainties from Planck constraints. The light and dark dashed blue lines depict the cases $A_{\rm ISW}=1$ (standard expectation) and $A_{\rm ISW}=0$, respectively.}\label{fig:f3}
\end{figure}

The last piece of the several data analyses carried out here is presented in what follows. Previous ISW analysis, as those developed by the Planck collaboration~\cite{Planck:2015fcm}
have never explored the possibility of binned analyses (both in terms of the redshift $z$ and of the scale $k$) of the ISW effect amplitude. We propose here a redshift tomography study of the ISW amplitude to be able to detect any possible deviation from the standard cosmological picture at the so-called dark ages, unreachable via other CMB observables. We have therefore considered the above mentioned 12-bin model. Figure~\ref{fig:f3} shows the results from such a redshift-bin ISW effect extraction as a function of the redshift $z$. Since the ISW amplitudes at $z=0$, 1300 and 1500 are unconstrained, we only illustrate the reconstructed values for the remaining ten redshift nodes. Table~\ref{tab:tab1} shows the reconstructed values of $A_{\rm ISW}$ for different redshift bins. In the SM, we depict the results for $32$ redshift bins, see Tab.~\ref{tab:tab32}.
\textit{Notice the $2\sigma$ deviation from the standard expectation $A_{\rm ISW} = 1$ at $z\simeq 500$, which is one of the most important results from our analysis.} A value of $A_{\rm ISW} > 1$ is equivalent to have a more intensive time variation of the gravitational potential during the dark ages. Therefore, this $2 \sigma$ signal around $z=500$ can be interpreted as: (i) a signal beyond either the standard General Relativity framework or the canonical dark matter fluid (as the dark energy component is irrelevant at these high redshifts); (ii) a signal beyond the standard ISW effect, and, finally, (iii) systematic effects which have been unaccounted for. The $\gtrsim 1\sigma$ significance of $A_{\rm ISW} < 1$ at $z=300$ and 700 may imply a possible oscillation correlation with $A_{\rm ISW} > 1$ at $z=500$. However, we do not find any oscillating mode in the ISW amplitude (see the SM).

\begin{table}[!t]
\renewcommand\arraystretch{1.5}
    \begin{center}
    \label{tab:tab1}
    \caption{Mean values of the ISW amplitudes with $1\sigma$ uncertainties at each redshift node in 12-bin model.}
    \setlength{\tabcolsep}{6mm}{   
    \begin{tabular}{l|c}
    \hline
    \hline
    Redshift & $A_{\rm ISW}$ \\
    \hline 
  10&  {\boldmath$A_{10}         $} = $0.71^{+0.34}_{-0.27}      $\\

30&{\boldmath$A_{30}         $} =$1.85^{+0.67}_{-1.80}       $\\

50&{\boldmath$A_{50}         $} = $1.76^{+0.60}_{-1.70}       $\\

80&{\boldmath$A_{80}         $} = $1.29^{+0.49}_{-1.30}       $\\

150&{\boldmath$A_{150}        $} = $1.05^{+0.41}_{-0.91}      $\\

300&{\boldmath$A_{300}        $} = $0.58^{+0.25}_{-0.44}      $\\

500&{\boldmath$A_{500}        $} = $1.63^{+0.41}_{-0.28}      $\\

700&{\boldmath$A_{700}        $} = $0.59^{+0.23}_{-0.49}      $\\

900&{\boldmath$A_{900}        $} = $0.97^{+0.39}_{-0.33}      $\\

1100&{\boldmath$A_{1100}       $} =$1.31\pm 0.49              $\\
    
    \hline
    \hline
    \end{tabular}}
    \end{center}
\end{table}

\textit{Discussions and conclusions.} CMB photons can help in the measurement of the gravitational potential and its evolution along the universe's history via the ISW effect. While previous studies in the literature have studied this effect, no research to date has completely studied the redshift and the scale dependence of the ISW effect. We propose here a novel tomographic method to extract useful physical information on the era of dark ages in the universe, otherwise inaccessible via any other cosmological probe. Considering one single ISW amplitude $A_{\rm ISW}$, the current Planck CMB data are able to measure it with $37\sigma$ significance. If, instead, we make use of ACT or SPT CMB data, we find a $\sim 6\sigma$ and $4\sigma$ detection of the ISW effect, respectively. We have firstly divided the ISW effect into an early ($A_{\rm{eISW}}$) and a late ($A_{\rm{lISW}}$) contribution, finding that Planck CMB data can provide a joint measurement of both parameters, ensuring a $38\sigma $ and a $2\sigma$ extraction of the former and latter, respectively. More importantly, this detection is robust against both model parameterizations and parameter degeneracies. 
For instance, the ISW detection and its significance do not depend on the lensing amplitude $A_{\rm L}$, even if these two parameters are extracted from the very same gravitational potential signatures. Leaving $A_{\rm L}$ as a free parameter, we find that $A_{\rm{eISW}}$ and $A_{\rm{lISW}}$
can be extracted with a significance of 
 $38\sigma$ and $2\sigma$, respectively. 
 In addition, the value of $A_{\rm L}$ is found to be perfectly consistent with $A_{\rm L}=1$,  $A_{\rm L} = 1.065^{+0.046}_{-0.060}$, and therefore the so-called lensing anomaly is diluted within this scenario. These results are stable against changes in the parameterization details, such as the transition redshift $z_t$ dividing the early and late ISW contributions. If one considers $z_t$ to be a free parameter, we find a $20\sigma$ and a $3\sigma$ detection of the early and late ISW effects. 

The novelty of our study also resides on the exploration of the redshift and the scale dependence of the ISW effect. We find that the ISW effect is well-consistent with the $\Lambda$CDM prediction on large scales (albeit this is only present in the range of scales between $10^{-4}h$~Mpc$^{-1}$ and $1h$~Mpc$^{-1}$) and we also find that all the ISW amplitudes for different $k$ nodes are perfectly compatible with the standard value $A_{\rm ISW}=1$. Indeed, a value of $A_{\rm ISW}>1$ would point to a stronger time variation of the gravitational potential during the dark ages. Our tomographic redshift analysis shows a $2\sigma$ deviation of the ISW amplitude ($A_{\rm ISW} >1$) at redshift $z=500$, and indicates that the ISW effect provides a very promising approach to explore the terra incognita at redshifts $z\sim \mathcal{O}(100)$ by probing the gravitational potential and its evolution. Possible physical solutions to explain this deviation may be modified gravity, dark radiation (for instance, an extra interaction between dark photons and the gravitational potential) or non-standard dark matter theories. 
Albeit it is premature to argue a new cosmological tension pointing to physics beyond the $\Lambda$CDM model under the General Relativity framework, future CMB cosmological measurements could unravel the origin of this departure from the canonical cosmological model.

\textit{Acknowledgements.}
This work has been supported by the Spanish MCIN/AEI/10.13039/501100011033 grants PID2020-113644GB-I00 and by the European ITN project HIDDeN (H2020-MSCA-ITN-2019/860881-HIDDeN) and SE project ASYMMETRY (HORIZON-MSCA-2021-SE-01/101086085-ASYMMETRY) and well as by the Generalitat Valenciana grants PROMETEO/2019/083 and CIPROM/2022/69.
DW is supported by the CDEIGENT Project of Consejo Superior de Investigaciones Científicas (CSIC). OM acknowledges the financial support from the MCIU with funding from the European Union NextGenerationEU (PRTR-C17.I01) and Generalitat Valenciana (ASFAE/2022/020).

\bibliography{isw}

\clearpage

\section{Supplementary Material}

\beginsupplement

\subsection{The amplitude of the ISW effect} 

\begin{figure*}
	\centering
	\includegraphics[scale=0.4]{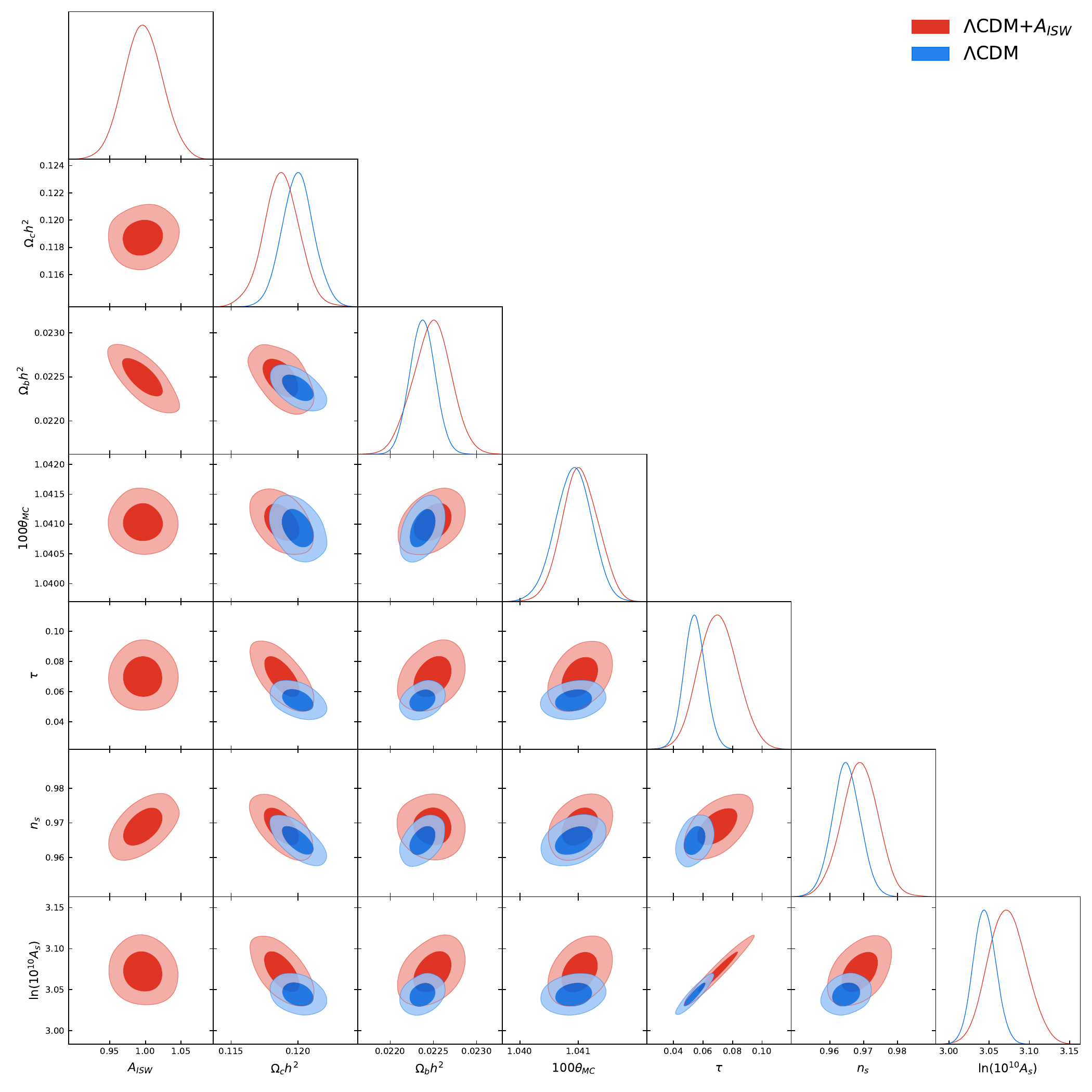}
	\caption{One-dimensional probability posterior distributions and two dimensional $68\%$ and $95\%$~CL allowed contours from Planck temperature, polarization and lensing observations for the most relevant cosmological parameters for the $\Lambda$CDM model with $A_{\rm ISW}=1$ and for a model in which $A_{\rm ISW}$ is a free parameter which rescales the overall amplitude of the ISW effect.}\label{fig:f4}
	
\end{figure*}

\begin{figure*}
	\centering
	\includegraphics[scale=0.45]{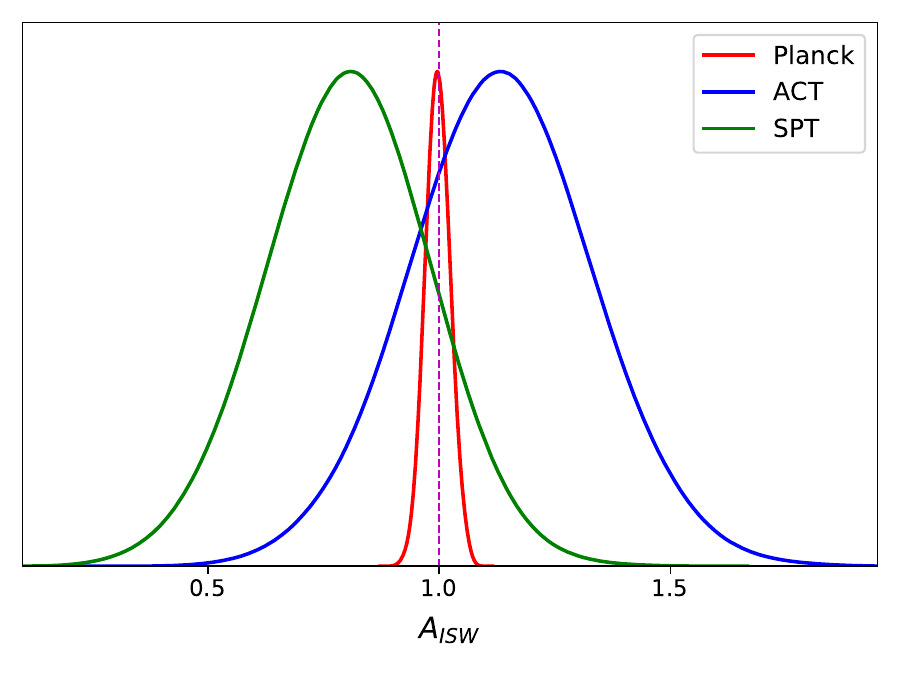}
        \includegraphics[scale=0.45]{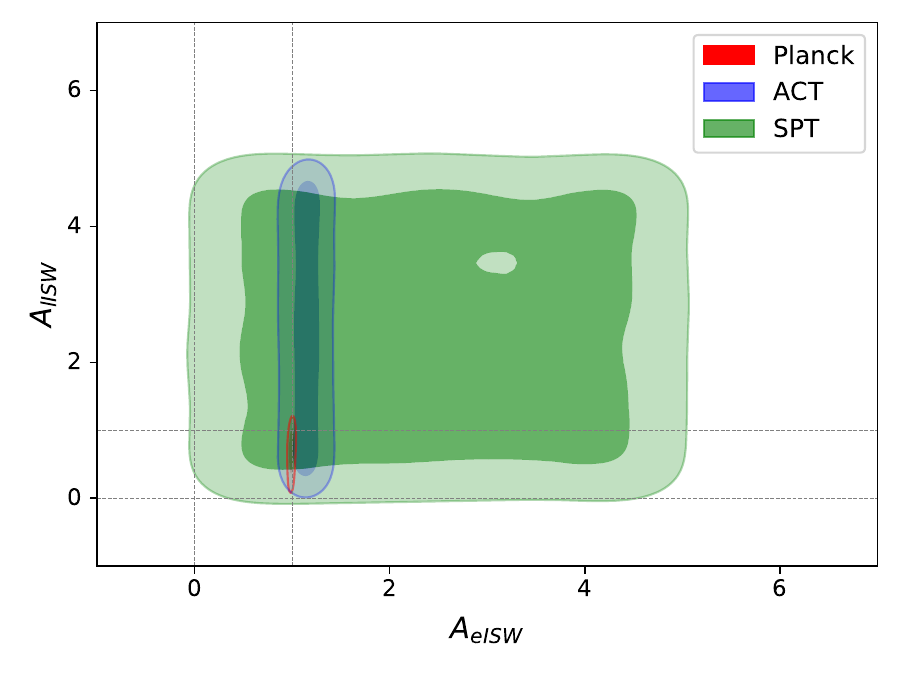}
	\caption{{\it left.} One-dimensional posterior distributions of the global ISW amplitude $A_{\rm ISW}$ from Planck, ACT and SPT CMB data, respectively. The dashed line denotes the standard expectation $A_{\rm ISW}=1$. {\it Right.} Two-dimensional posterior distributions of ($A_{\rm eISW}$, $A_{\rm lISW}$) from Planck, ACT and SPT CMB data, respectively. We fix $z_t=30$ for two-parameter model. }\label{fig:a1}
	
\end{figure*}

The ISW effect constitutes a secondary source of CMB anisotropies and it is represented by Eq.~(\ref{eq:isw}). Figure~\ref{fig:f4} depicts the one-dimensional probability distributions and the two-dimensional allowed contours for the different cosmological parameters when a parameter $A_{\rm ISW}$ rescales Eq.~(\ref{eq:isw}), governing the amplitude of the ISW effect. The observations exploited to derive the allowed regions are CMB temperature, polarization and lensing data from the Planck 2018 final release~\cite{Planck:2018vyg,Planck:2019nip,Planck:2018lbu}. As a comparison, we depict the minimal $\Lambda$CDM results assuming  $A_{\rm ISW}=1$.
The value we obtain from Planck data is $A_{\rm ISW} = 0.997\pm 0.027$ at $68\%$~CL, which implies a $\sim 37\sigma$ detection of the global ISW effect. We are also interested in the constraining power of two ground-based CMB detector ACT and SPT on $A_{\rm ISW}$ and the corresponding result is shown in the left panel of Fig. \ref{fig:a1}. ACT and SPT give the constraining values $A_{\rm ISW} = 1.13 \pm 0.20$ and $0.80 \pm 0.18$, indicating a $6\sigma$ and a $4\sigma$ detection of the ISW effect, respectively. Interestingly, SPT-3G presents a beyond $1\sigma$ hint of a smaller ISW amplitude and ACT prefers a larger ISW effect. 
Notice that, in the case in which $A_{\rm ISW}$ is a free parameter, the allowed range for the scalar spectral index $n_s$, for the reionization optical depth $\tau$ and for the amplitude of the scalar perturbations $A_s$ is much larger, due to the degeneracies among these three parameters. 
A larger value of $A_{\rm ISW}$ will increase the amplitude of the ISW effect around the first peak. Such an enhancement should be accompanied by an increase of both the tilt of the power spectrum $n_s$ (that will increase the slope of the power spectrum of the primordial scalar power spectrum $P_s(k)$, raising the right side relative to the left side) and the amplitude of the scalar fluctuations $A_s$. Since $A_s$ is strongly degenerate with $e^{-2\tau}$, a larger value of $A_s$ automatically implies a smaller value of the reionization optical depth. Consequently, the mean values for the these three parameters ($\tau$, $n_s$ and $A_s$) are larger than in the standard $\Lambda$CDM picture. The correlation with $A_{\rm ISW}$ and the matter content of the universe is also positive. As we will shortly see, an increase of $\Omega_c h^2$ leads to a larger value of $z_{eq}$, making the suppression of the ISW effect larger, and therefore requiring an increase of the ISW amplitude $A_{\rm ISW}$ to compensate for such an effect. Notice also the strong anti-correlation between  $A_{\rm ISW}$ and the baryon energy density: a larger value of $\Omega_b h^2$ will suppress the second peak compared with the first and third ones, requiring therefore the ISW effect to be smaller in order to also reduce the first peak, where the \emph{early} ISW contribution is maximal, as we shall shortly show. 

\subsection{Early and late ISW effects}

In the fully matter dominated period, the gravitational potentials are almost constant in time and therefore, following Eq.~(\ref{eq:isw}), the ISW effect will be very small. 
Right after recombination there is still a radiation component present in the universe, and assuming a vanishing anisotropic stress and evaluating the Bessel function at recombination ($\eta_r$), the early ISW effect reads as:

\begin{equation}
 \Theta_\ell (k)\simeq 2 j_\ell (\eta_r) \left[\Phi (k, \eta_m)-\Phi(k,\eta_r)\right]~,
\end{equation}
where the gravitational potential is evaluated in the matter dominated regime $\eta_m$. Notice that the early ISW effect ~\cite{Cabass:2015xfa} adds in phase with the primary anisotropy, increasing the height of the first acoustic peaks, with an emphasis on the first one, due to the fact that the main contribution of the ISW effect is at scales $k\sim 1/\eta_e$, i.e. around the first acoustic peak. In addition, the early ISW effect will be suppressed by the square of the radiation-to matter ratio $\propto [(1+z_r)/(1+z_{eq})]^2$, i.e. a larger (smaller) matter component will result into a smaller (larger) ISW amplitude due to the larger (smaller) value of $z_r$. This is the reason for the degeneracy between $\Omega_c h^2$ and $A_{\rm ISW}$. On the other hand, a larger baryon energy density will lead to a higher recombination redshift, leading to a larger ISW amplitude and to an anti-correlation between $\Omega_b h^2$ and $A_{\rm ISW}$. The enhancement factor of the ISW effect amplitude makes this effect an excellent observable to identify extra relativistic particles present at recombination.

Conversely, the late ISW effect is only present when dark energy starts to be important in the universe's evolution,  being this effect only measurable at large scales. The time-dependent gravitational potentials are precisely responsible for both the development of galaxy clusters and for weak lensing distortions. The Planck collaboration reported a $4\sigma$ detection of the late ISW effect cross-correlating temperature anisotropies with both the lensing potential and galaxy number counts.

In the right panel of Fig. \ref{fig:a1}, we show the the allowed contours in the parameter space ($A_{\rm eISW}$, $A_{\rm lISW}$) from Planck, ACT and SPT, respectively. It is easy to see that Planck gives the strongest constraints on the parameter space, SPT can not provide any constraint on these two parameters, and ACT only provides a constraint on the amplitude of the early ISW effect.

\subsection{ISW Parameterizations and analysis}
\label{sec:param}
\begin{figure*}
	\centering
	\includegraphics[scale=0.4]
{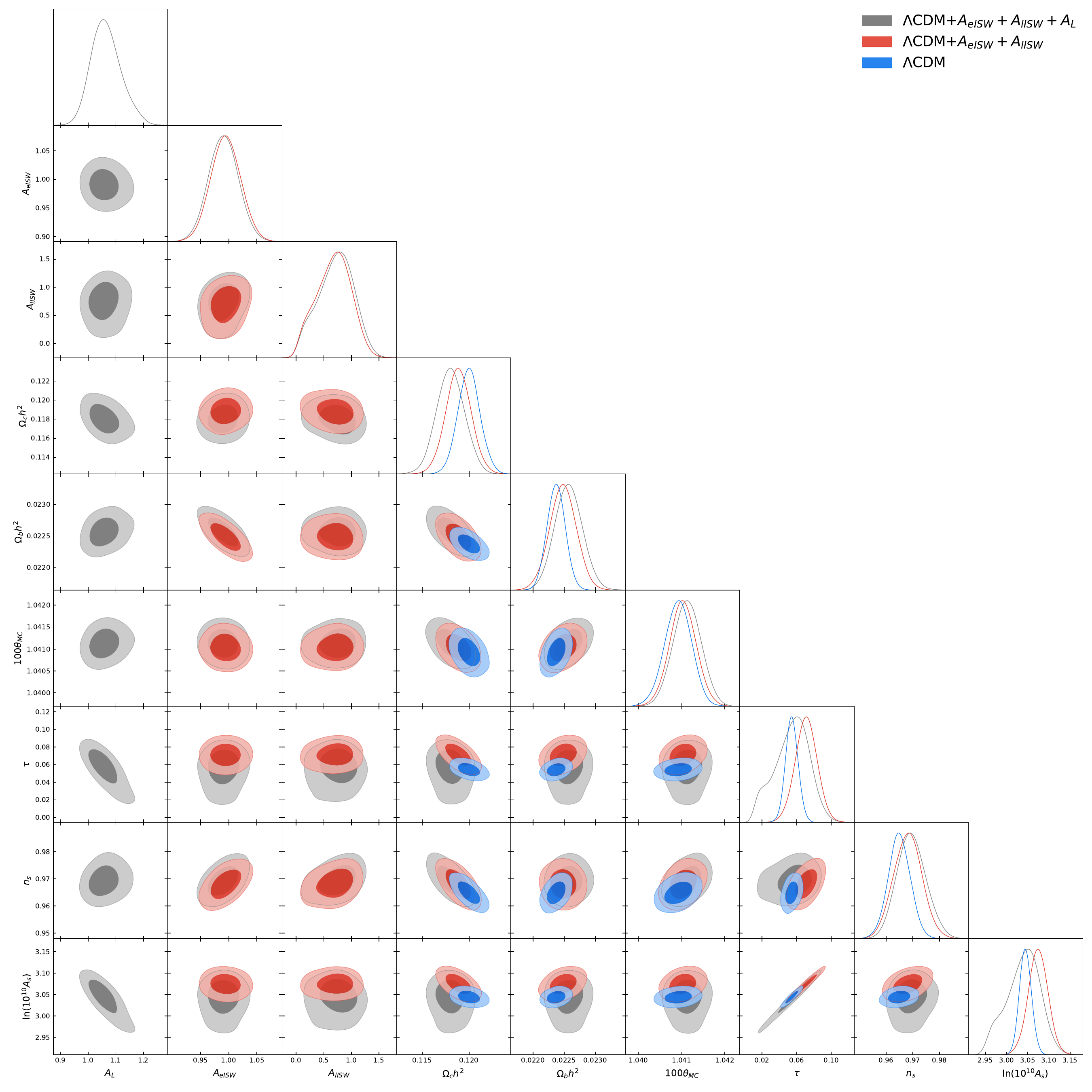}
	\caption{One-dimensional probability posterior distributions and two dimensional $68\%$ and $95\%$~CL allowed contours from Planck temperature, polarization and lensing observations for the most relevant cosmological parameters for the $\Lambda$CDM model with $A_{\rm ISW}=1$, for a model in which $A_{\rm e ISW}$ and $A_{\rm l ISW}$ are free parameters, rescaling the overall amplitude of the ISW effect, and for a model in which, together with the two former free parameters, the lensing amplitude $A_{\rm L}$ is also a free parameter.}
 \label{fig:f5}
\end{figure*}

Following Ref.~\cite{Cabass:2015xfa}, we explore here a two-redshift bin parameterization  of the ISW
amplitude in terms of two parameters $A_{\rm{eISW}}$ and $A_{\rm{lISW}}$, which rescale the contribution at early and late times of the ISW effect, multiplying Eq.~(\ref{eq:isw}) with an amplitude $A_{\rm{eISW}}$ if $z>z_t=30$ and with an amplitude  $A_{\rm{lISW}}$ if $z<z_t=30$. Figure ~\ref{fig:f5} depicts the one-dimensional probability distributions and the two-dimensional allowed contours for the different cosmological parameters when the two  parameters $A_{\rm{eISW}}$ and $A_{\rm{lISW}}$, governing the amplitude of the ISW effect, 
rescale Eq.~(\ref{eq:isw}) at early and late times respectively. The observations exploited to derive the allowed regions are CMB temperature, polarization and lensing data from the Planck 2018 release~\cite{Planck:2018vyg,Planck:2019nip,Planck:2018lbu}. As a comparison, we also depict the $\Lambda$CDM results, assuming  $A_{\rm ISW}=1$. The values we obtain are $A_{\rm{eISW}} = 0.995\pm 0.026$ and 
$A_{\rm{lISW}} = 0.67^{+0.35}_{-0.28}$, implying a $37\sigma$ and a $2\sigma$ detection of the early and late ISW effects respectively from CMB data only.

Notice that not only the mean value and errors on the early ISW effect and those previously found for the parameterization in terms of one single parameter $A_{\rm ISW}$ are very similar: also the main degeneracies among the parameters look very close to those previously described when only a single ISW amplitude was introduced in the $\Lambda$CDM parameter content. 

Figure~\ref{fig:f5} also illustrates  the case in which the lensing amplitude $A_{\rm L}$~\cite{Calabrese:2008rt} is  a free parameter in the analysis, that we shall comment on below. 
CMB anisotropies get blurred due to gravitational lensing: photons from different directions are mixed and the acoustic peaks at large multipoles are smoothed. The amount of lensing is a precise prediction of the $\Lambda$CDM model: the consistency of the model can be checked by artificially increasing lensing by a higher amplitude factor $A_{\rm{L}}$~\cite{Calabrese:2008rt}  (\emph{a priori}, an unphysical parameter). If $\Lambda$CDM consistently describes all CMB data, observations should prefer $A_{\rm{L}}=1$. The so-called lensing anomaly~\cite{Planck:2018vyg,Motloch:2018pjy} is due to the fact that Planck CMB temperature and polarization prefers $A_\mathrm{L} > 1$ at 3$\sigma$. Notice that both the one-dimensional and two-dimensional probability distributions are very similar for the cases $A_{\rm{L}}=1$ and a freely varying lensing amplitude: current CMB observations are perfectly able to disentangle between the lensing and the ISW effects. In addition, the lensing anomaly dilutes when the ISW amplitude is rescaled. Figure~\ref{fig:f5} shows very mild degeneracies between $A_{\rm L}$ and the ISW amplitudes $A_{\rm{eISW}}$ and 
$A_{\rm{lISW}}$. There is a minor anti-correlation between $A_{\rm L}$ and the early ISW effect, due to their different location in the CMB angular power spectrum: while the early ISW effect is maximal around the first peak, the lensing amplitude smooths the higher $\ell$ CMB acoustic peaks : a  
$20\%$ change in $A_{\rm L}$ suppresses the fourth and higher peaks by $\sim 0.5\%$, raising troughs by $\sim1\%$. 

 \begin{table}
\renewcommand\arraystretch{1.5}
    \begin{center}
    \caption{Mean values of the ISW amplitude with $1\sigma$ uncertainties at each redshift node in the 32-bin model.}
    \setlength{\tabcolsep}{6mm}{
    \label{tab:tab32}
    \begin{tabular}{l|c}
    \hline
   \hline
    Redshift & $A_{\rm ISW}$ \\
    \hline 
10&{\boldmath$A_{10}         $} = $0.48^{+0.63}_{-0.24}      $\\

50&{\boldmath$A_{50}         $} = $0.2\pm 1.5                $\\

100&{\boldmath$A_{100}        $} = $0.3^{+3.0}_{-2.4}         $\\

150&{\boldmath$A_{150}        $} = $1.1^{+2.6}_{-1.8}         $\\

200&{\boldmath$A_{200}        $} = $0.7^{+3.6}_{-1.9}         $\\

250&{\boldmath$A_{250}        $} = $1.7^{+2.8}_{-1.3}         $\\

300&{\boldmath$A_{300}        $} = $0.5^{+3.6}_{-2.2}         $\\

350&{\boldmath$A_{350}        $} =$0.6^{+2.9}_{-2.3}         $\\

400&{\boldmath$A_{400}        $} = $0.1\pm 2.7                $\\

450&{\boldmath$A_{450}        $} = $0.7^{+3.2}_{-2.0}         $\\

500&{\boldmath$A_{500}        $} = $1.3^{+3.3}_{-1.5}         $\\

550&{\boldmath$A_{550}        $} = $2.38^{+2.40}_{-0.85}       $\\

600&{\boldmath$A_{600}        $} = $1.9^{+2.9}_{-1.1}         $\\

650&{\boldmath$A_{650}        $} = $1.8^{+2.8}_{-1.4}         $\\

700&{\boldmath$A_{700}        $} = $-0.6^{+2.0}_{-3.5}        $\\

750&{\boldmath$A_{750}        $} = $-0.8^{+2.2}_{-2.7}        $\\

800&{\boldmath$A_{800}        $} = $0.5^{+3.7}_{-2.2}         $\\

850&{\boldmath$A_{850}        $} = $1.9^{+2.9}_{-1.0}         $\\

900&{\boldmath$A_{900}        $} = $1.7^{+3.1}_{-1.2}         $\\

950&{\boldmath$A_{950}        $} = $1.3^{+3.1}_{-1.6}         $\\

1000&{\boldmath$A_{1000}       $} = $0.2^{+3.0}_{-2.5}         $\\

1050&{\boldmath$A_{1050}       $} = $1.1^{+2.7}_{-1.8}         $\\

1100&{\boldmath$A_{1100}       $} = $0.6^{+3.6}_{-2.0}         $\\

1150&{\boldmath$A_{1150}       $} = $1.8^{+3.0}_{-1.1}         $\\

1200&{\boldmath$A_{1200}       $} = $0.2^{+3.6}_{-2.5}         $\\

1250&{\boldmath$A_{1250}       $} = $0.1\pm 2.9                $\\

1300&{\boldmath$A_{1300}       $} = $-0.2^{+2.4}_{-3.8}        $\\

1350&{\boldmath$A_{1350}       $} = $0.4\pm 2.9                $\\

1400&{\boldmath$A_{1400}       $} = $0.1\pm 2.8                $\\

1450&{\boldmath$A_{1450}       $} = $0.1\pm 2.7                $\\

1500&{\boldmath$A_{1500}       $} = $-0.2^{+2.3}_{-3.9}        $\\

1550&{\boldmath$A_{1550}       $} = $0.2^{+3.8}_{-2.4}         $\\

    \hline
    \hline
    \end{tabular}}
    \end{center}
\end{table} 

\begin{figure*}
	\centering
	\includegraphics[scale=0.4]{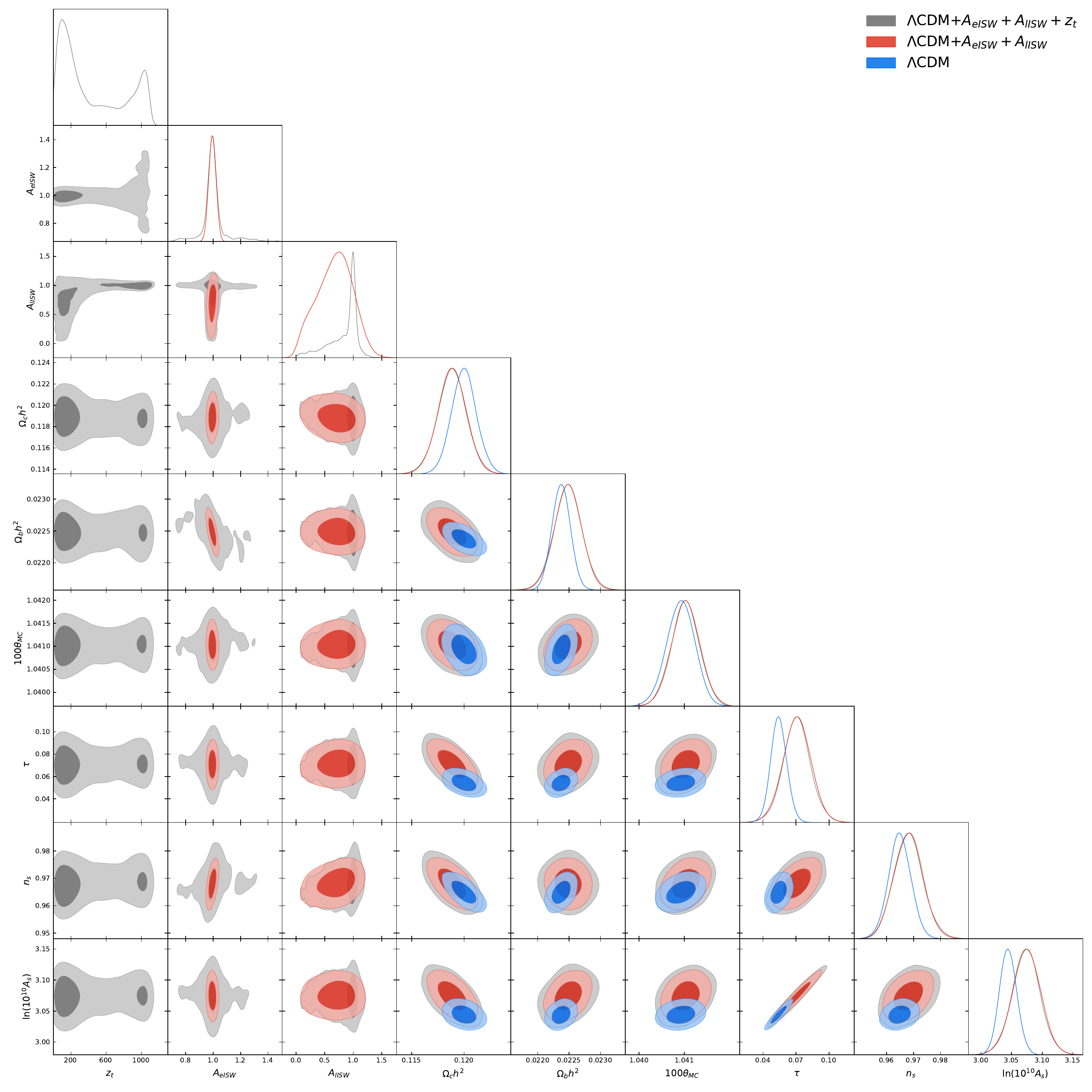}
	\caption{One-dimensional probability posterior distributions and two dimensional $68\%$ and $95\%$~CL allowed contours from Planck temperature, polarization and lensing observations for the most relevant cosmological parameters for the $\Lambda$CDM model with $A_{\rm ISW}=1$, for a model in which $A_{\rm e ISW}$ (ISW amplitude for $z>z_t$) and $A_{\rm l ISW}$ (ISW amplitude for $z<z_t$) are free parameters together with the transition redshift $z_t$, and for a model in which, together with the three former free parameters, the lensing amplitude $A_{\rm L}$ is also a freely varying parameter.}\label{fig:f6}
\end{figure*}

Figure~\ref{fig:f6} depicts the equivalent to Fig.~\ref{fig:f5} but assuming a freely varying $z_t$. Notice the bimodal one-dimensional probability distribution for $z_t$, showing two distinct peaks, located at $z_t\simeq 1000$ and $z_t\simeq100$, where the early and late ISW effects are expected to be located. We obtain the constraint $z_{t} = 434^{+700}_{-400}$ from Planck CMB data, which is inconsistent with $z_t=30$ at beyond the $1\sigma$ level. The addition of $z_t$ as a free parameter diminishes the statistical significance of the sensitivity to the early ISW effect, albeit the former is always above the $20\sigma$ level. Interestingly, the one-dimension posterior distribution of $A_{\rm lISW}$ has a narrow peak, while the sensitivity to the late ISW effect remains unchanged at $2\sigma$ level. This means that a varying transition redshift $z_{t}$ can capture the information of the late ISW effect more accurately than fixing $z_t=30$. To a large extent, it can describe better the effect of dark energy on the gravitational potential and consequently the contribution of dark energy to CMB at large scales. 

\begin{figure*}
	\centering
	\includegraphics[scale=0.4]{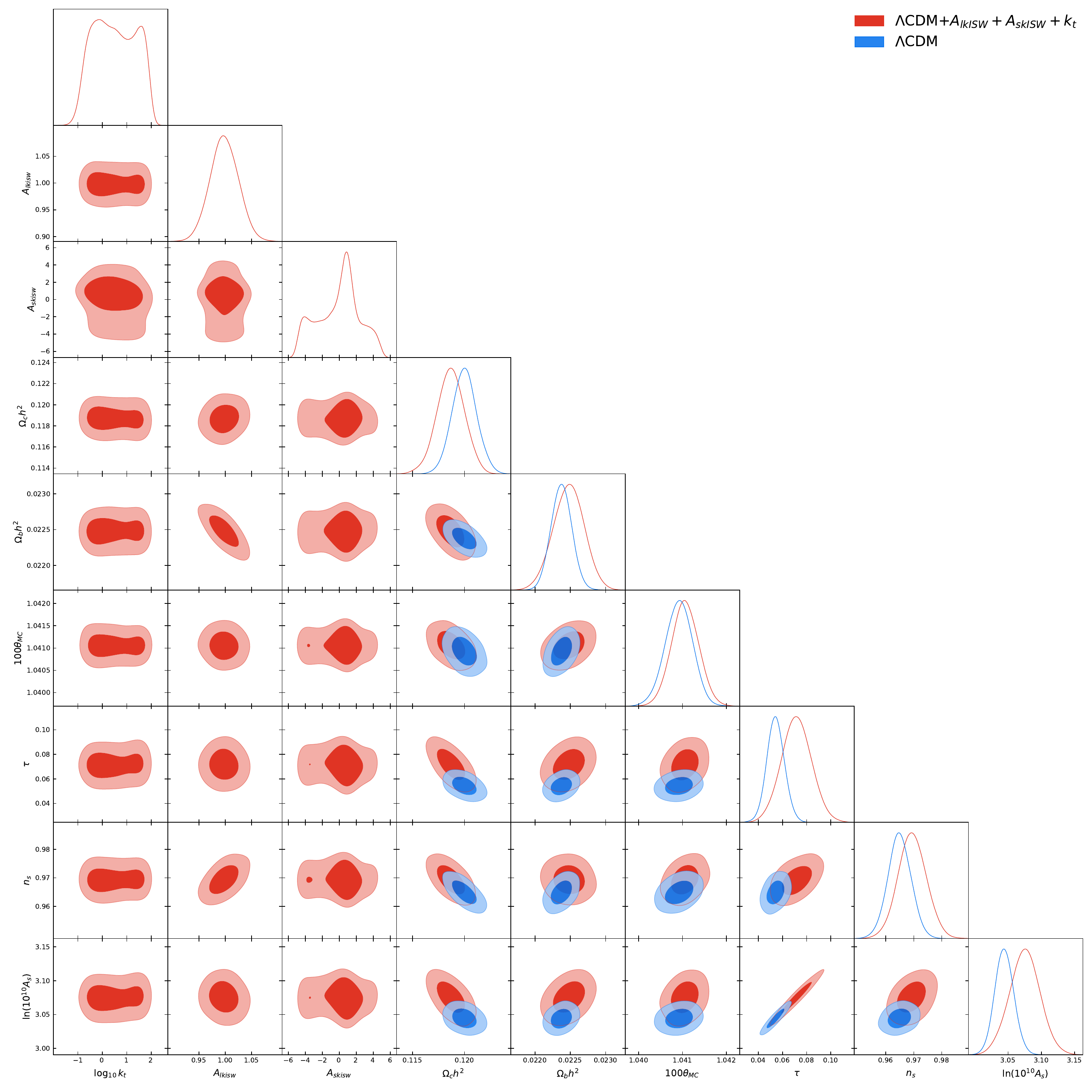}
	\caption{One-dimensional probability posterior distributions and two dimensional $68\%$ and $95\%$~CL allowed contours from Planck temperature, polarization and lensing observations for the most relevant cosmological parameters for the $\Lambda$CDM model with $A_{\rm ISW}=1$, and for a model in which $A_{\rm lkISW}$ (ISW amplitude for $k<k_t$) and $A_{\rm skISW}$ (ISW amplitude for $k>k_t$) are free parameters together with a possible transition scale $k_t$.}\label{fig:f7}
\end{figure*}

\begin{figure*}
	\centering
	\includegraphics[scale=0.3]{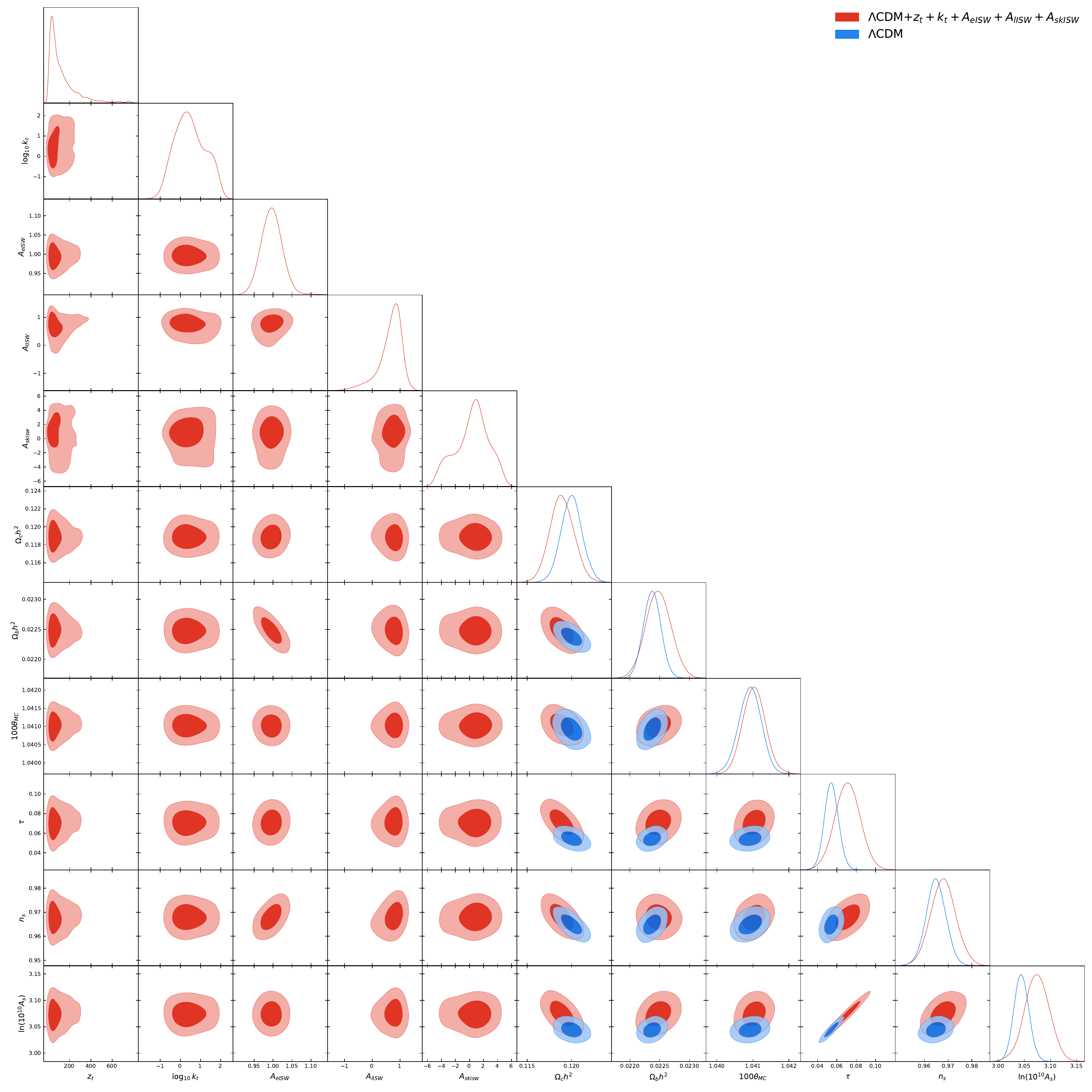}
	\caption{One-dimensional probability posterior distributions and two dimensional $68\%$ and $95\%$~CL allowed contours from Planck temperature, polarization and lensing observations for the most relevant cosmological parameters for the $\Lambda$CDM model with $A_{\rm ISW}=1$, and for a model in which $A_{\rm e ISW}$ (ISW amplitude for $z>z_t$), $A_{\rm l ISW}$ (ISW amplitude for $z<z_t$) and $A_{\rm skISW}$ (ISW amplitude for $k>k_t$) are free parameters together with both the transition redshift $z_t$ and the transition scale $k_t$.}\label{fig:f8}
\end{figure*}

Figures~\ref{fig:f7} and \ref{fig:f8} illustrate the results of two distinct analysis to explore the possibility of a non-negligible transition scale $k_t$.  In one case, we have fixed $z_t=30$, while in the other we have leave $z_t$ as a free parameter. For the former case, we obtain the $1\sigma$ constraint ${\rm log}_{10}\,k_t = 0.52\pm 0.85$ in $k$ range, indicating that the ISW effect mainly works on large scales ${\rm log}_{10}\ (k_t < -0.33$ at $1\sigma$ CL), which is very compatible with theoretical predictions. The mean values and $1\sigma$ uncertainties of the ISW amplitudes are $A_{\rm lkISW} = 0.998\pm 0.027$ and $A_{\rm skISW} = -0.1^{+2.5}_{-3.1}$, implying a $37\sigma$ detection of the large scale ISW effect and a null contribution from small scales, respectively. The rest of the contours look exactly the same than those obtained in the absence of a free $k_t$ parameter. 
For the latter case, we adopt the multiple tomographic method to characterize the ISW effect. Firstly, we divide the ISW amplitude into a large scale part and a small scale part via a transition scale $k_t$ and one amplitude $A_{\rm skISW}$. Then, we divide the large scale ISW effect into two separate parts, i.e., the early and late ISW effects via the transition redshift $z_t$ and two amplitudes $A_{\rm eISW}$ and $A_{\rm lISW}$. Hence, this model actually has five parameters. Confronting this model against  Planck CMB data, we obtain the 
$1\sigma$ constraints $z_t=127^{+20}_{-100}$ and $k_{t} = 0.50^{+0.64}_{-0.97}$, which gives a simultaneous determination of the transition redshift and scale. Similar to the free $k_t$ case, we do not find small scale ISW effect according to the constraint $A_{\rm skISW} = 0.4^{+3.0}_{-1.9}$. The constraining values $A_{\rm eISW} = 0.996\pm 0.028$ and $A_{\rm lISW} = 0.65^{+0.47}_{-0.22}$ indicate that the signal of the large scale ISW effect can be measured with a significance of $36\sigma$ in the early case and of $2\sigma$ in the late case.
It is easy to see that this model enhances the significance of early ISW effect relative to $20\sigma$ detection in the free $z_t$ case, because we only include the gravitational potential information at relatively large scales for early and late ISW effects in this scenario. Furthermore, the enlargement of model parameter space leads to the reduced significance of the early ISW effect relative to the $38\sigma$ detection in the $z_t=30$ case, since $z_t=127^{+20}_{-100}$ is consistent with $z_t=30$ at $1\sigma$ level.

\subsection{Tomographic-redshift ISW analysis}
\label{sec:tomoz} 

A redshift tomography study of the ISW amplitude should be able to detect any possible deviation from the standard cosmological picture during the era of dark ages, unobservable by any other probe, except for 21cm experiments and only in their low-$z$ region.

\begin{figure*}
	\centering
	\includegraphics[scale=0.18]{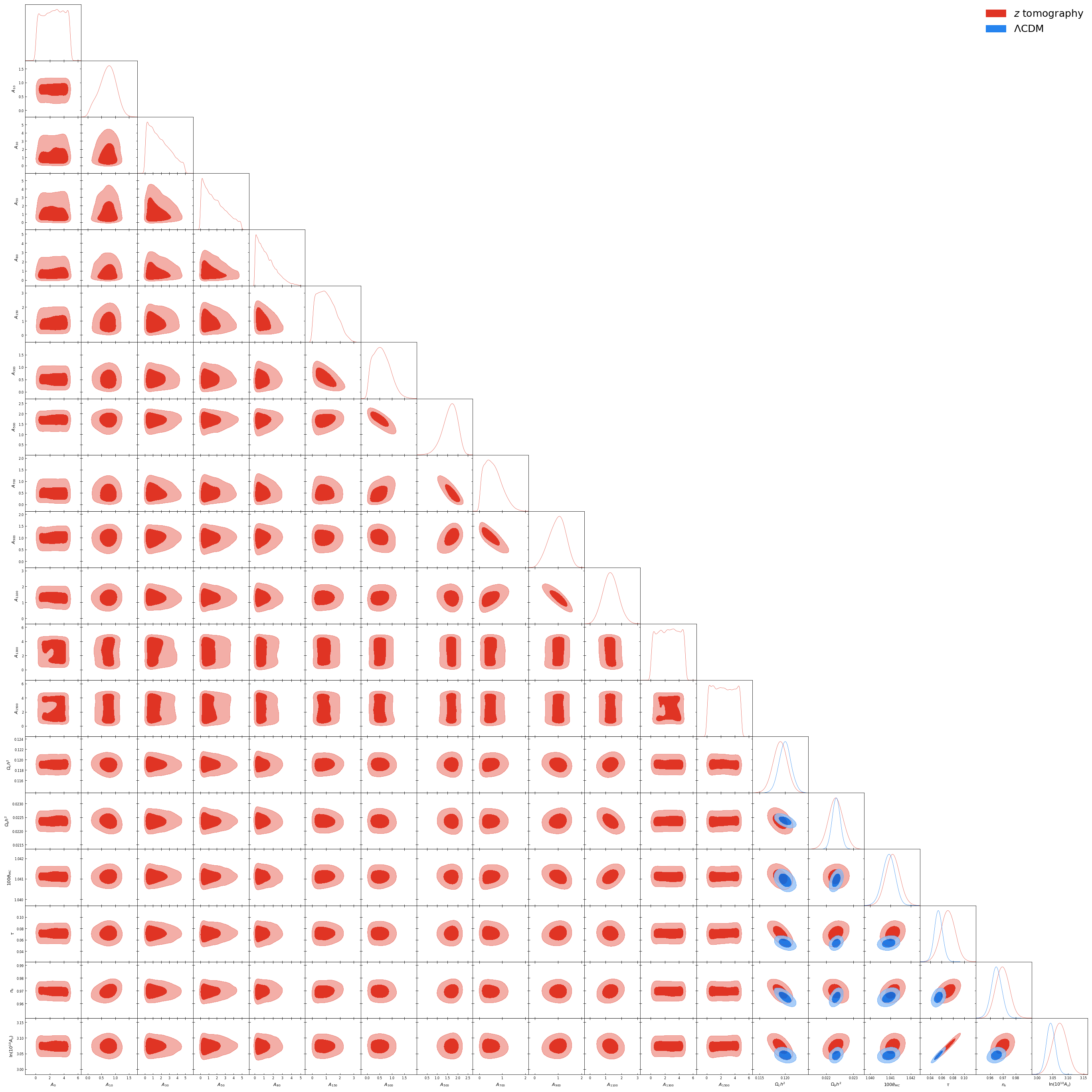}
	\caption{One-dimensional probability posterior distributions and two dimensional $68\%$ and $95\%$~CL allowed contours from Planck temperature, polarization and lensing observations for the most relevant cosmological parameters for the $\Lambda$CDM model with $A_{\rm ISW}=1$ and for a model with 12 redshift nodes each of them associated to a different $A_{\rm ISW}$, see text for details.}\label{fig:f9}
\end{figure*}

\begin{figure*}
	\centering
	\includegraphics[scale=0.09]{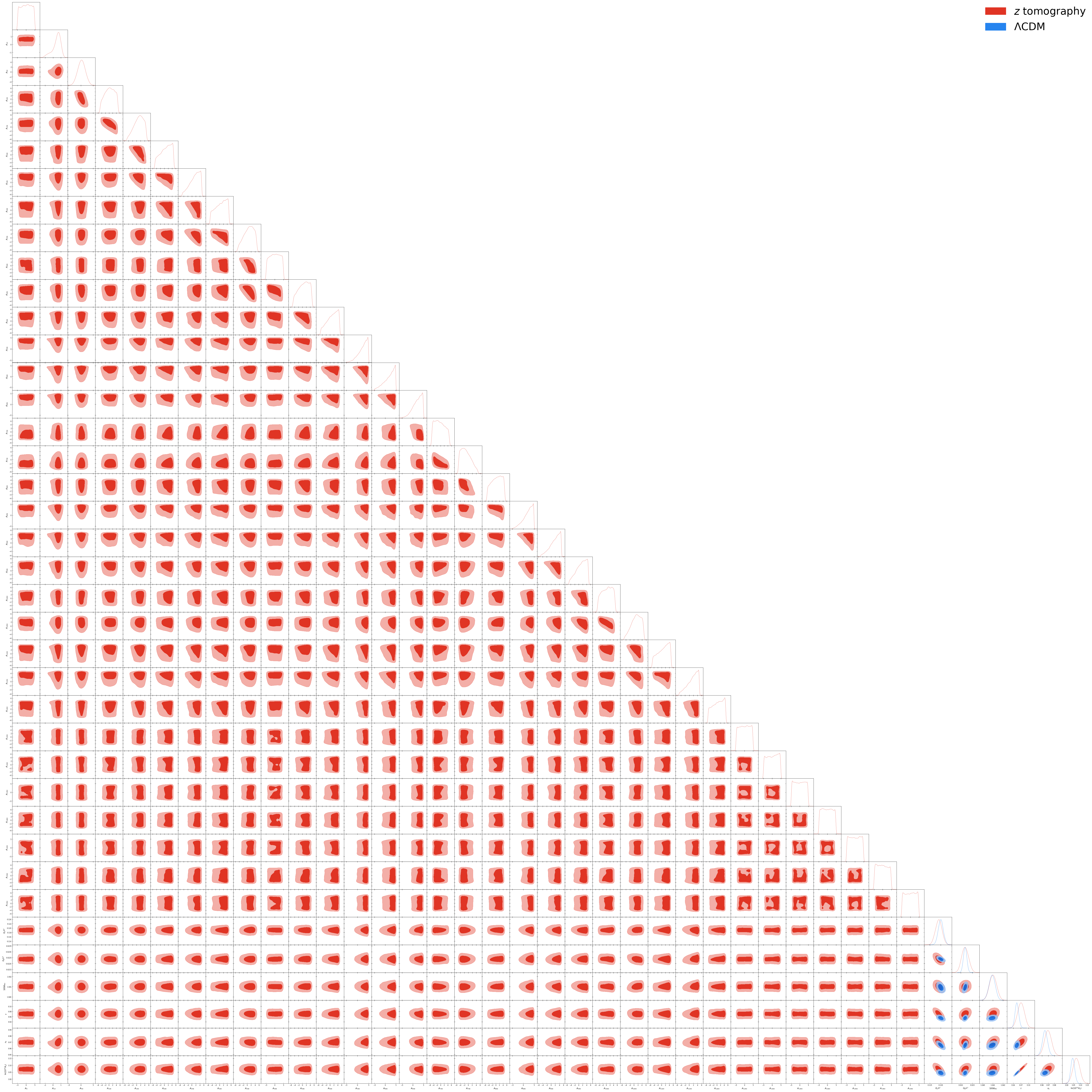}
	\caption{One-dimensional probability posterior distributions and two dimensional $68\%$ and $95\%$~CL allowed contours from Planck temperature, polarization and lensing observations for the most relevant cosmological parameters for the $\Lambda$CDM model with $A_{\rm ISW}=1$ and for a model with 32 redshift nodes each of them associated to a different $A_{\rm ISW}$, see text for details.}\label{fig:f10}
\end{figure*}

We have presented in the main manuscript the constraints on the ISW amplitudes assuming twelve redshift nodes in the range $z\in [0,\,1600]$. The marginalized posterior distributions of this 12-bin model are shown in Fig.~\ref{fig:f9}. 
In order to study carefully the excess of the ISW amplitude around $z=500$ and verify the robustness of our results, we shall illustrate the case with thirty-two nodes in the very same redshift range here. Table~\ref{tab:tab32} depicts the mean values together with their $1\sigma$ uncertainties on the ISW amplitudes and Fig.~\ref{fig:f10} shows the marginalized $1\sigma$ and $2\sigma$ posterior contours.

\textit{We still observe a $\sim2\sigma$ deviation from the standard expectation $A_{\rm ISW} = 1$, at $z=550$, $A_{550}=2.38^{+2.40}_{-0.85}$, which establishes the robustness of our findings, which are shown to be independent of the redshift bin size. Therefore, this $\sim 2 \sigma$ signal around $z\simeq 550$ is stable and, if not due to systematic effects, could be signaling new physics during the so-called dark ages of the universe.}

\subsection{Tomographic-scale ISW analysis}
\label{sec:tomok}

\begin{figure*}
	\centering
	\includegraphics[scale=0.25]{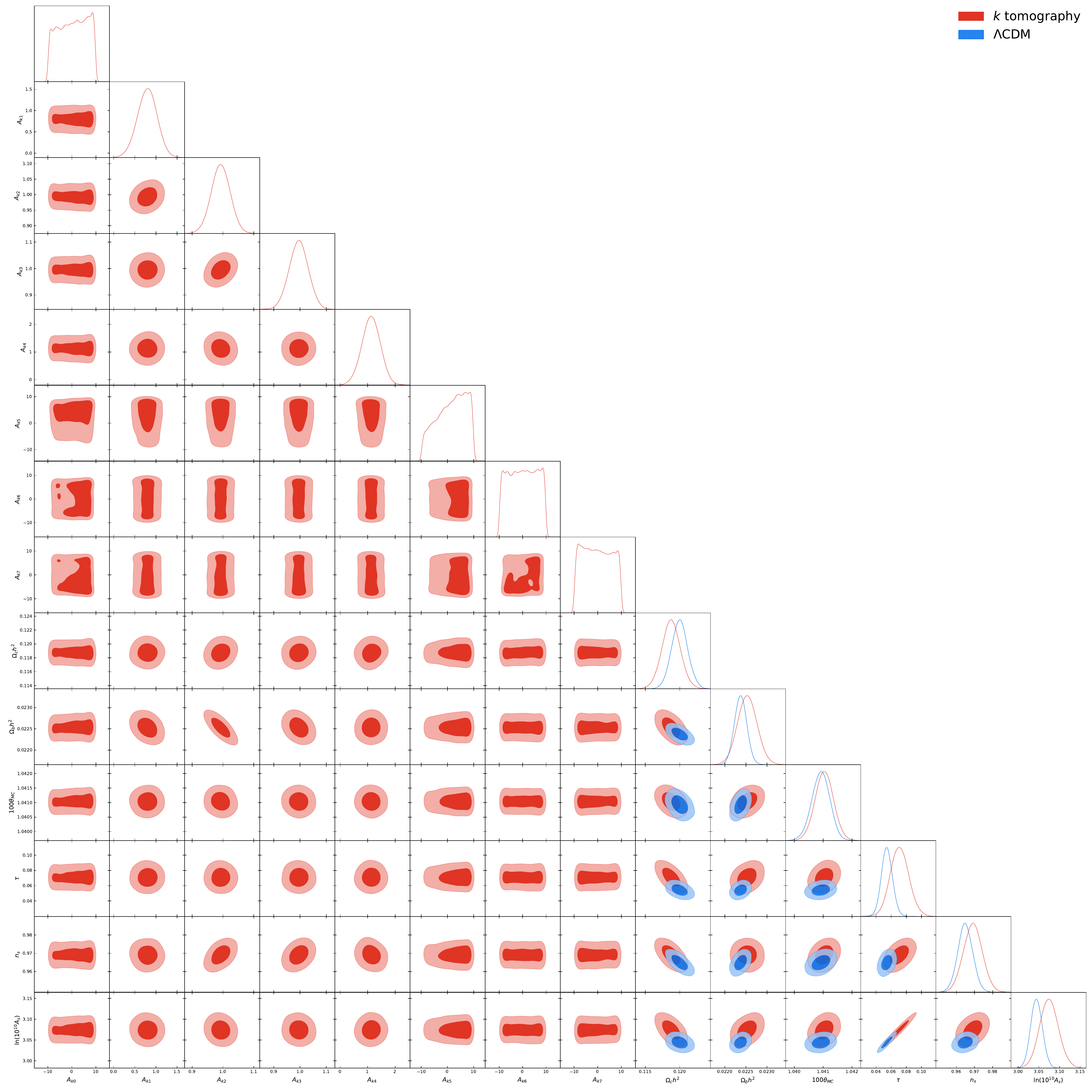}
	\caption{One-dimensional probability posterior distributions and two dimensional $68\%$ and $95\%$~CL allowed contours from Planck temperature, polarization and lensing observations for the most relevant cosmological parameters for the $\Lambda$CDM model with $A_{\rm ISW}=1$ and for a model with eight nodes in logarithmic $k$ space each of them associated to a different $A_{\rm ISW}$, see text for details.}\label{fig:f11}
\end{figure*}

We have also performed an analysis with eight different nodes in $k$ space, each of them associated to a different ISW amplitude. Namely, if $k \ge 1h$~Mpc$^{-1}$, $A_{\rm ISW }=A_{k0}$, if 
$1h$~Mpc$^{-1}$ $> k \ge 0.1h$~Mpc$^{-1}$, $A_{\rm ISW }=A_{k1}$, if $0.1 $~Mpc$^{-1}$ $> k \ge 0.01h$~Mpc$^{-1}$, $A_{\rm ISW }=A_{k2}$, if $0.01h$~Mpc$^{-1}$ $> k \ge 0.001h$~Mpc$^{-1}$, $A_{\rm ISW}=A_{k3}$, if $0.001h$~Mpc$^{-1}$ $> k \ge 10^{-4}h$~Mpc$^{-1}$, $A_{\rm ISW}=A_{k4}$, if $ 10^{-4}h $~Mpc$^{-1}$ $> k \ge 10^{-5}h$~Mpc$^{-1}$, $A_{\rm ISW }=A_{k5}$,
if $10^{-5}h$~Mpc$^{-1}$ $> k \ge 10^{-6}h$~Mpc$^{-1}$, $A_{\rm ISW }=A_{k6}$, and, finally, if $ 10^{-6}h $~Mpc$^{-1}$ $> k$, $A_{\rm ISW }=A_{k7}$. 
Figure~\ref{fig:f11} shows the marginalized $1\sigma$ and $2\sigma$ constraints from Planck temperature, polarization and lensing data on these eight amplitudes together with the $\Lambda$CDM parameters. 
Notice that the amplitudes $A_{k0}$, $A_{k5}$, $A_{k6}$ and $A_{k7}$ are compatible with zero, meaning a turn-on and a turn-off in $k$ scale of the ISW effect.  The remaining amplitudes, $A_{k1}=0.79\pm 0.23$, $A_{k2}=0.993\pm 0.031$, $A_{k3}=0.995\pm 0.036$ and $A_{k4}=1.13\pm 0.34$ are basically consistent with the standard expectation $A_{\rm ISW}=1$ at $1\sigma$ level and show almost identical degeneracies with the remaining cosmological parameters, clearly stating the fact that the ISW effect is well compatible with $\Lambda$CDM prediction in the $k$-space window of ($0.1$, $10^{-4}$)~Mpc$^{-1}$. This is the main conclusion from our tomographic analysis in $k$-space. Furthermore, it is interesting that $A_{k1}$ presents a small $\sim1\sigma$ discrepancy with $A_{\rm ISW}=1$. This implies that there may exist a stronger time variation of the gravitational potential around cluster scales. The four well constrained amplitudes $A_{k1}$, $A_{k2}$, $A_{k3}$ and $A_{k4}$ imply $3\sigma$, $32\sigma$, $28\sigma$ and $3\sigma$ detections of the ISW effect, respectively. One can easily conclude that the Planck collaboration mainly detected the ISW effect in the range $k\in[0.001, 0.1]h$ Mpc$^{-1}$, which is generally beyond the cluster scale.

\subsection{Multiple tomographic method}

The multiple tomographic method (MTM) we use contains the following two classes: (i) general MTM; (ii) special MTM. The former one is the combination of $z_t$-only and $k_t$ only, while the latter one denotes the 5-parameter extension to $\Lambda$CDM $z_t+k_t+A_{\rm skISW}+A_{\rm eISW}+A_{\rm lISW}$, namely:

$$ 
{\rm MTM}\Longrightarrow\,\left\{
\begin{aligned}
&z_t,\,\,\,k_t                                                   & \Longrightarrow {\rm General \quad MTM}                                                     \\
&z_t+k_t                                                    & \Longrightarrow {\rm Special \quad MTM}                                                      \\
\end{aligned}
\right. \label{eq:MTM}
$$

The general MTM can help in obtaining more complete information about the ISW effect in each dimension ($z_t$ or $k_t$). This means that we consider the full scale information of the gravitational potentials when dividing the ISW effect into early and late parts via $z_t$. Its shortcoming is that one can not determine simultaneously the transition redshift and the transition scale. Interestingly, the special MTM can help in finding out the single transition spacetime point $(z_t, k_t)$. However, it neglects the gravitational potential information at some medium and small scales. In this analysis, the combined use of the two MTMs can provide a complete characterization of the ISW effect. If we must take only one MTM method to analyse the ISW effect, one would prefer the general MTM one because it will contain more physical information input and it is more conservative.  

\subsection{Stability analysis}

\begin{figure*}
	\centering
	\includegraphics[scale=0.3]{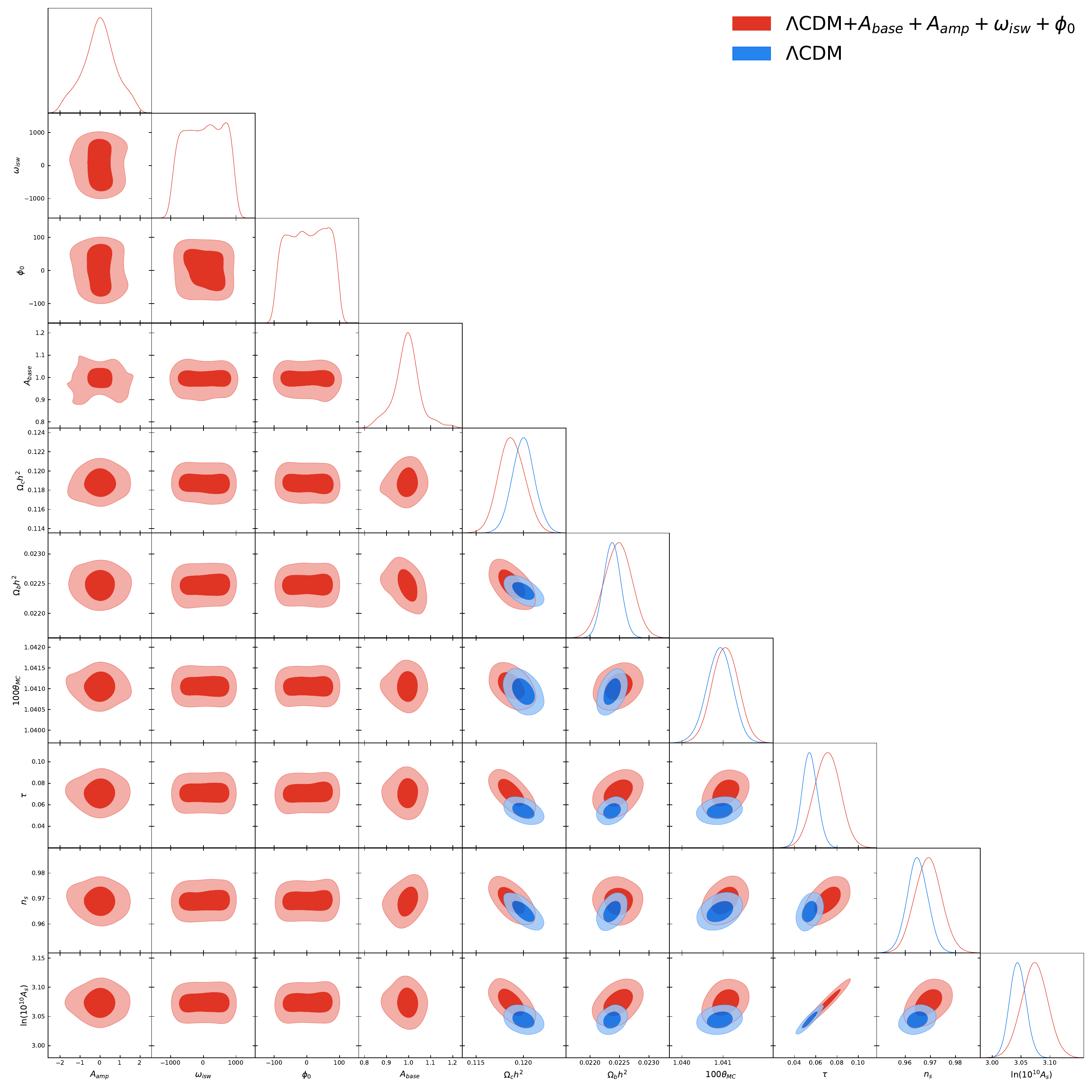}
	\caption{One-dimensional probability posterior distributions and two dimensional $68\%$ and $95\%$~CL allowed contours from Planck temperature, polarization and lensing observations for the most relevant cosmological parameters for the $\Lambda$CDM model with $A_{\rm ISW}=1$ and for an oscillating ISW amplitude model, see text for details.}\label{fig:f12}
\end{figure*}

As described above, we observe a possible oscillating mode within the 12-redshift-bin model in the range $z\in[200,\,800]$. This implies that there may exist an oscillating ISW amplitude over redshift. To solve this puzzle, we propose an oscillating ISW amplitude $A_{\rm ISW}(z)=A_{base}+A_{amp}[\sin(\omega_{isw}z)+\phi_0]$ and constrain it with Planck CMB data (see Fig.~\ref{fig:f12}). We obtain the $1\sigma$ constraints $A_{base} = 0.993^{+0.029}_{-0.022}$ and $A_{amp} = -0.03\pm 0.85$, implying that there is no significant oscillation of the ISW amplitude through the cosmic history and a $34\sigma$ evidence of the ISW effect is observed in the context of this model. At the very same time, current data can not give any constraint on the frequency and the initial phase. The very same conclusion is reached within the 32-bin model. The result from this oscillating model is well consistent with that from the tomographic analysis.

\begin{table}
\renewcommand\arraystretch{1.5}
    \begin{center}
    \caption{Mean values of the amplitude of the gravitational potential with $1\sigma$ uncertainties at each redshift node in the 17-bin model.}
    \setlength{\tabcolsep}{6mm}{
    \label{tab:smt2}
    \begin{tabular}{l|c}
    \hline
   \hline
    Redshift & $A_{\rm \Phi}$ \\
    
\hline

0.5 & {\boldmath$A_{\Phi0.5}         $}=$1.08^{+0.100}_{-0.084}     $\\

1.5 & {\boldmath$A_{\Phi1.5}         $}=$1.00\pm 0.11              $\\

4 & {\boldmath$A_{\Phi4}         $}=$0.992\pm 0.051            $\\

8 & {\boldmath$A_{\Phi8}         $}=$1.055\pm 0.070            $\\

30 & {\boldmath$A_{\Phi30}         $}=$0.998\pm 0.023            $\\

100 & {\boldmath$A_{\Phi100}         $}=$1.001\pm 0.031            $\\

200 & {\boldmath$A_{\Phi200}         $}=$1.028\pm 0.091            $\\

300 & {\boldmath$A_{\Phi300}         $}=$0.97\pm 0.20              $\\

400 & {\boldmath$A_{\Phi400}         $}=$1.03\pm 0.36              $\\

500 & {\boldmath$A_{\Phi500}         $}=$0.83^{+0.38}_{-0.59}      $\\

600 & {\boldmath$A_{\Phi600}         $}=$1.53^{+0.64}_{-0.37}      $\\

700 & {\boldmath$A_{\Phi700}         $}=$0.52^{+0.15}_{-0.51}      $\\

800 & {\boldmath$A_{\Phi800}         $}=$1.07\pm 0.55              $\\

900 & {\boldmath$A_{\Phi900}         $}=$1.94^{+0.74}_{-1.90}       $\\

1000 & {\boldmath$A_{\Phi1000}         $}=$0.71\pm 0.32              $\\

1100 & {\boldmath$A_{\Phi1100}         $}=$1.18^{+0.21}_{-0.18}      $\\

1200 & {\boldmath$A_{\Phi1200}         $}=$0.76^{+0.22}_{-0.26}      $\\
    \hline
    \hline
    \end{tabular}}
    \end{center}
\end{table}

\begin{figure*}
	\centering
	\includegraphics[scale=0.5]{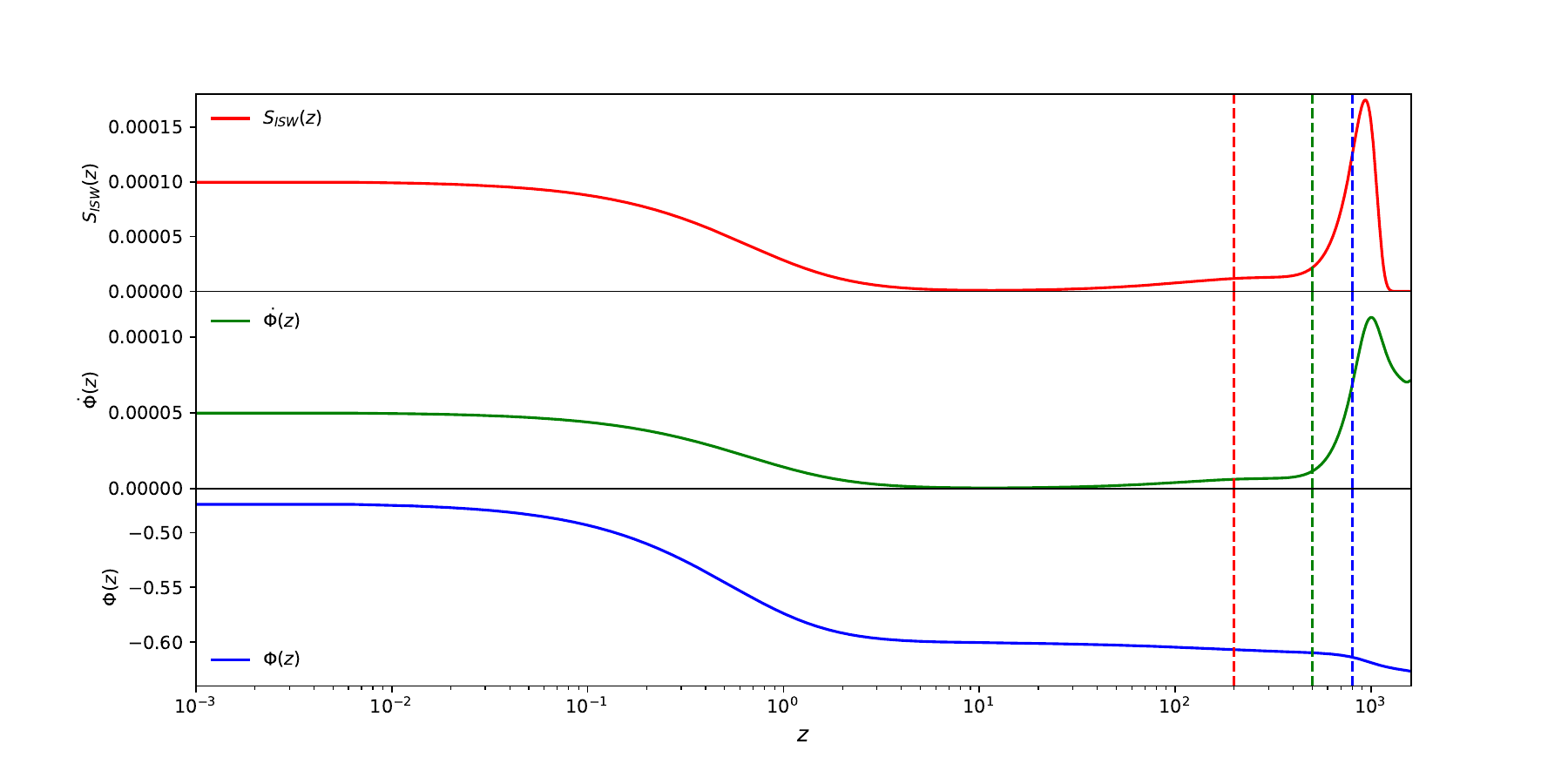}
	\caption{The CMB ISW source function $S_{\rm ISW}(z)$, time derivative of gravitational potential $\dot{\Phi}(z)$ and gravitational potential $\Phi(z)$ as a function of redshift $z$ at a given large scale $k=10^{-5}\,h$ Mpc$^{-1}$, respectively. The vertical dashed red, green and blue lines denote $z=200,\,500$, and 800, respectively. }\label{fig:f13}
\end{figure*}

\begin{figure*}[htbp]
\centering
\includegraphics[scale=0.42]
{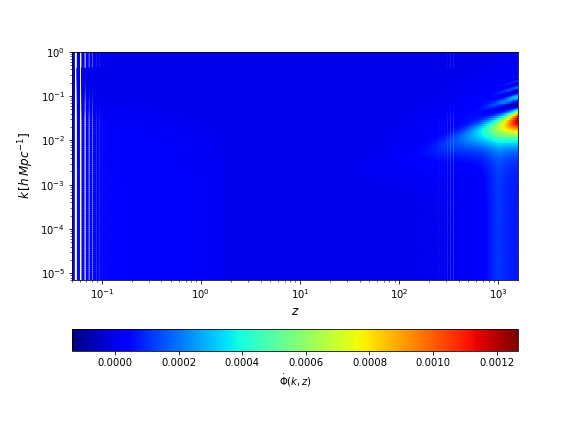} 
\includegraphics[scale=0.42]
{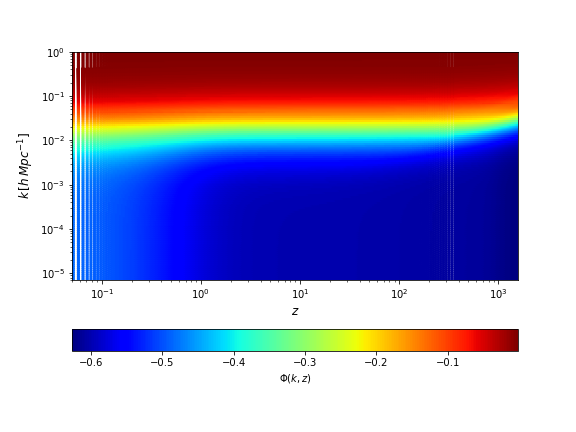}
 	\caption{{\it Left.} The time derivative of gravitational potential $\dot{\Phi}(k,\,z)$ as a function of redshift $z$ and scale $k$. {\it Right.} The gravitational potential $\Phi(k,\,z)$ as a function of redshift $z$ and scale $k$.}
   \label{fig:f14}
\end{figure*}

\begin{figure*}
	\centering
	\includegraphics[scale=0.4]{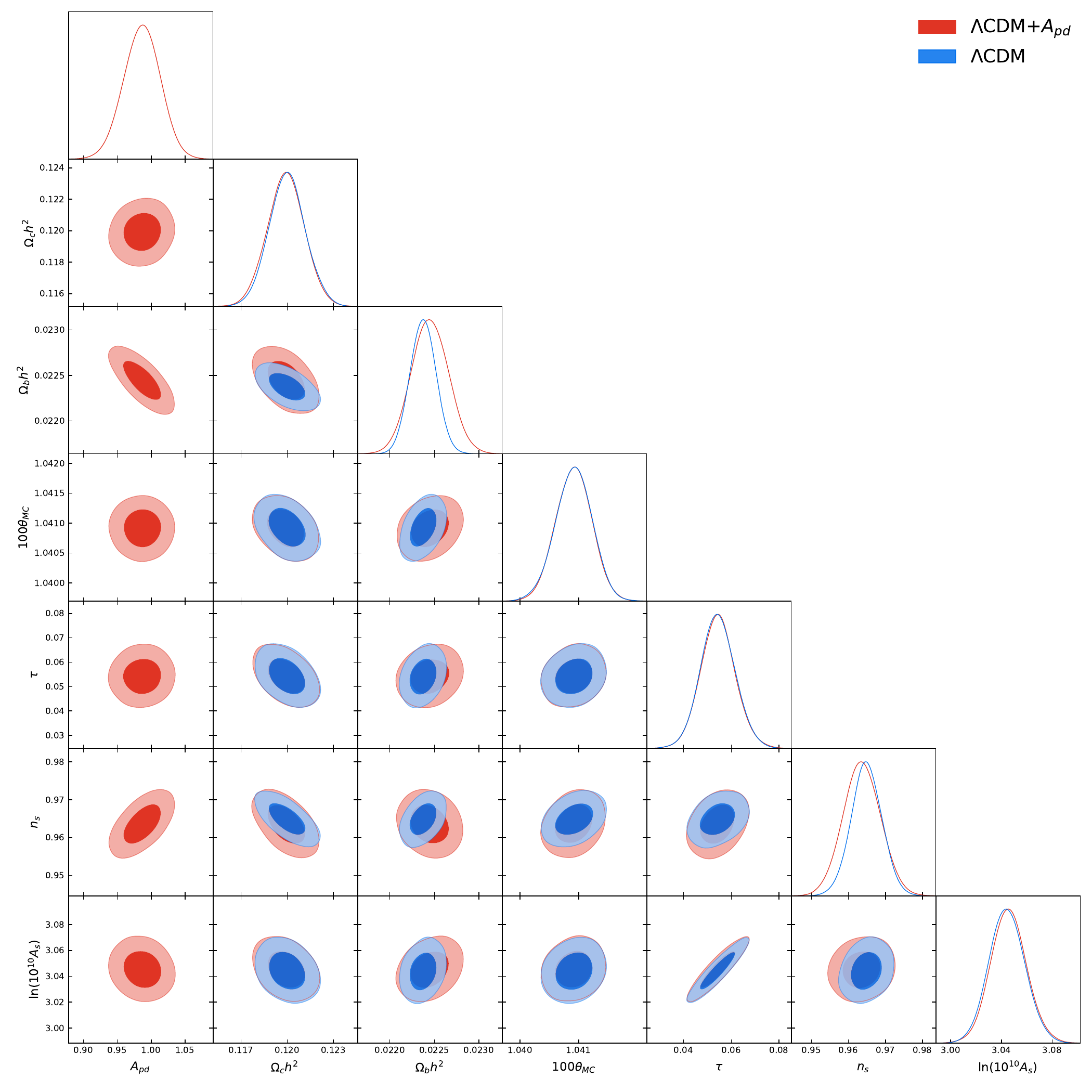}
	\caption{One-dimensional probability posterior distributions and two dimensional $68\%$ and $95\%$~CL allowed contours from Planck temperature, polarization and lensing observations for the most relevant cosmological parameters and for a model in which $A_{\rm pd}$ is a free parameter which rescales the overall amplitude of the time derivative of gravitational potential.}\label{fig:f15}
\end{figure*}

\begin{figure*}
	\centering
	\includegraphics[scale=0.4]{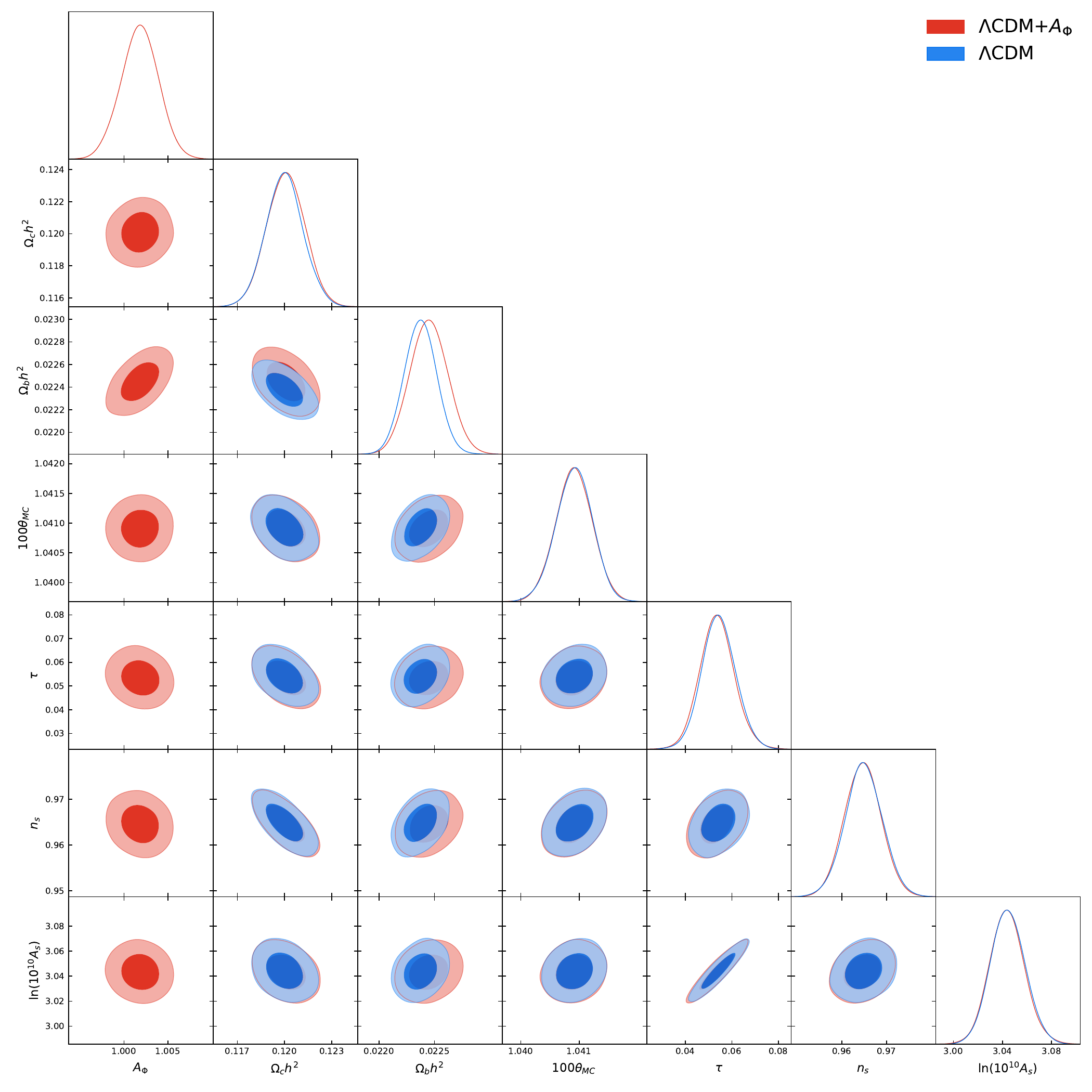}
	\caption{One-dimensional probability posterior distributions and two dimensional $68\%$ and $95\%$~CL allowed contours from Planck temperature, polarization and lensing observations for the most relevant cosmological parameters and for a model in which $A_{\rm {\Phi}}$ is a free parameter which rescales the overall amplitude of the gravitational potential.}\label{fig:f16}
\end{figure*}

A closely related issue to the ISW effect is to extract the gravitational potential and its time derivative based on current CMB data. By analyzing the potential details, we can find out the source responsible for the  $2\sigma$ deviation from the theoretical prediction $A_{\rm ISW}=1$. In Fig.~\ref{fig:f13}, as an example, we show the time derivative of gravitational potential $\dot{\Phi}(z)$ and the gravitational potential $\Phi(z)$ as a function of redshift $z$ at a given large scale $k=10^{-5}\,h$ Mpc$^{-1}$, respectively. One can easily find that the redshift gradient of the potential $\dot{\Phi}(z)$ traces well the evolution of $S_{\rm ISW}(z)$ during the cosmic expansion (see Fig.~\ref{fig:f2}), since these two quantities are different by only a factor of $e^{-\tau(z)}$. While the gravitational potential always decays starting from the recombination epoch, its slope also becomes larger and larger, although it varies very slowly at some stages such as within the $z\in[10,\,100]$ range. Note that even if the potential is very flat in $z\in[10,\,100]$, $S_{\rm ISW}(z)$ and $\dot{\Phi}(z)$ are not exactly zero. Actually, they are very small (e.g. $\mathcal{O}(-14)$) and close to zero due to the existence of neutrinos and a very small portion of radiation. Moreover, we observe that $S_{\rm ISW}(z)$ is clearly non-zero in the range $z\in[200,\,800]$, where we find the $2\sigma$ deviation from $\Lambda$CDM in the 12-bin model. Furthermore, we present the two-dimensional maps of the gravitational potential and its time derivative over redshift and scale in Fig.~\ref{fig:f14}. It is easy to see that $\dot{\Phi}(z)$ almost exhibits the same evolution as $S_{\rm ISW}(z)$, since its value is always close to a half of $S_{\rm ISW}(z)$. $\dot{\Phi}(z)$ has its maximal value around recombination at about $k=10^{-2}\,h$ Mpc$^{-1}$. As depicted in Fig.~\ref{fig:f13}, the gravitational potential decays over redshift. However, this decay is scale-dependent. Specifically, $\Phi(z)$ has a larger scale gradient at high redshifts (say, recombination epoch) and it decays with the increase of scale $k$. To compare with  the ISW amplitude models, we also assume the similar amplitudes for $\dot{\Phi}(z)$ and $\Phi(z)$, and then we confront them against CMB data, and the corresponding results are shown in Figs.~\ref{fig:f15} and \ref{fig:f16}. We obtain the tight $1\sigma$ constraints on the amplitude of $\dot{\Phi}(z)$, $A_{pd} = 0.986\pm 0.027$, and on  the amplitude of $\Phi(z)$, $A_{\Phi} = 1.0018\pm 0.0021$. This result is well consistent with the $\Lambda$CDM prediction and gives the same $37\sigma$ detection as the ISW effect (see the main text) and a $477\sigma$ evidence of the gravitational potential. This incredibly large significance ($477\sigma$) is very crucial to for modern cosmology, since the properties of gravitational potential $\Phi(z)$ govern the large scale structure formation of the universe.        

\begin{figure*}
	\centering
	\includegraphics[scale=0.18]{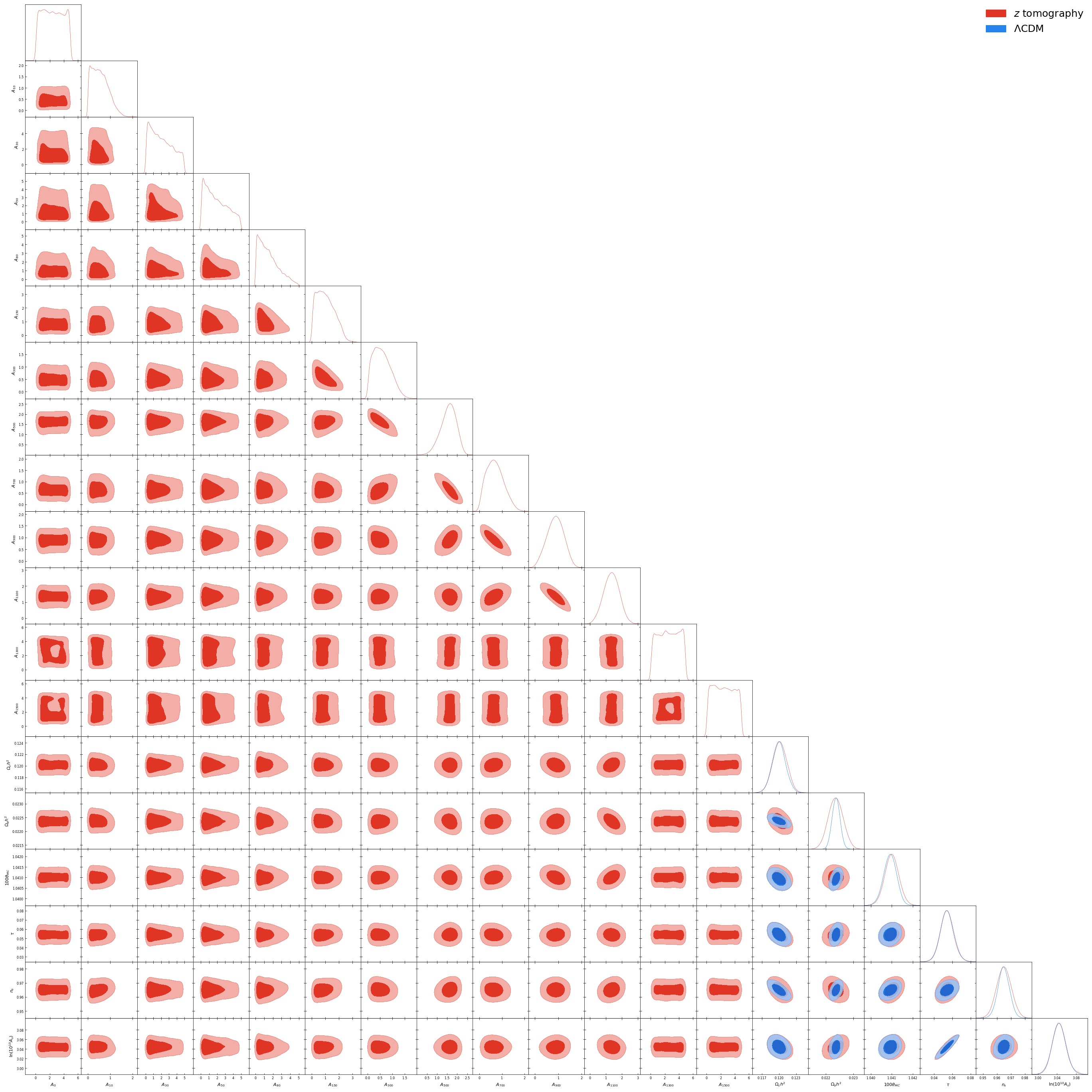}
	\caption{One-dimensional probability posterior distributions and two dimensional $68\%$ and $95\%$~CL allowed contours from Planck temperature, polarization and lensing observations for the most relevant cosmological parameters for the $\Lambda$CDM model and for a model with 13 redshift nodes each of them associated to a different $A_{\rm pd}$.}\label{fig:f17}
\end{figure*}

\begin{figure*}
	\centering
	\includegraphics[scale=0.15]{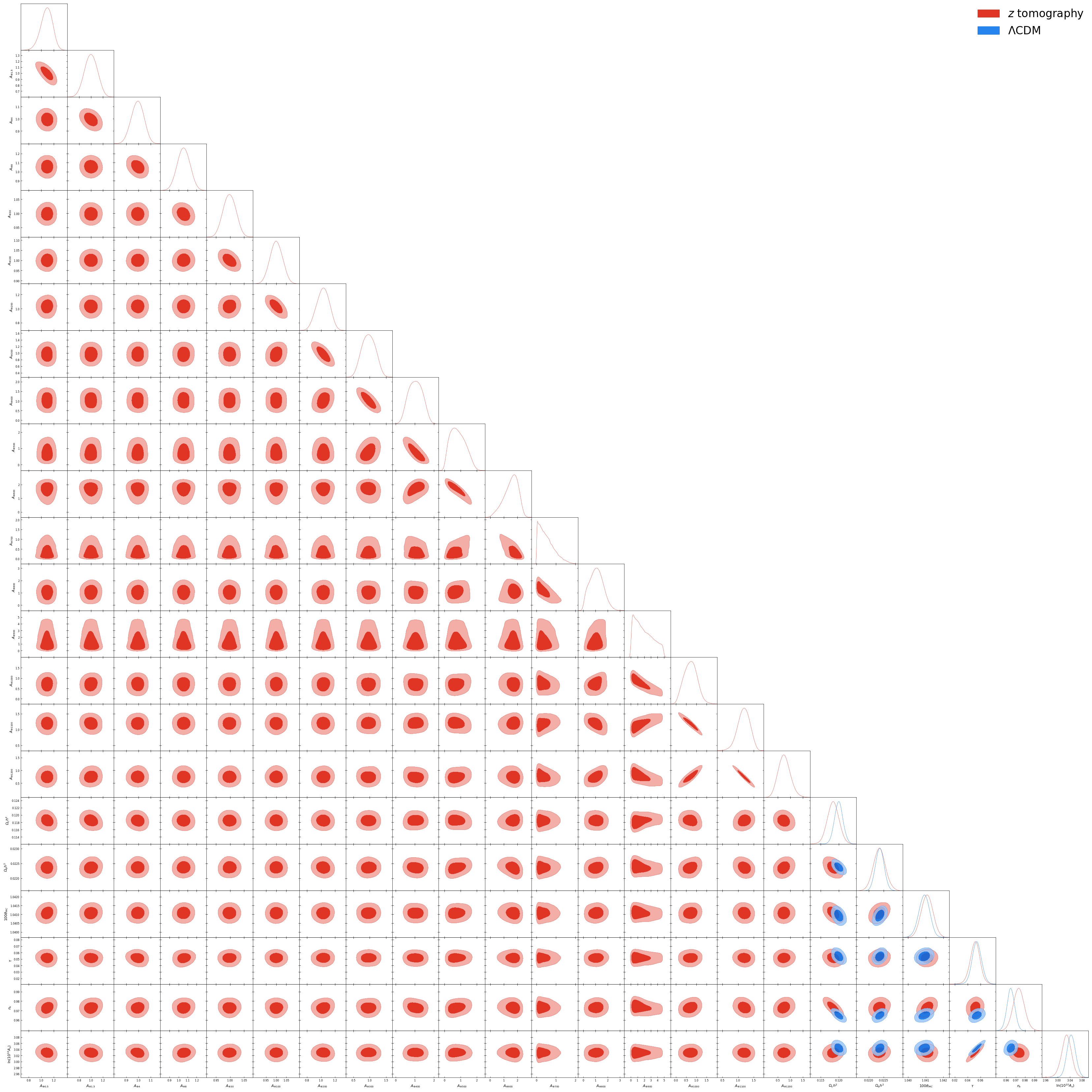}
	\caption{One-dimensional probability posterior distributions and two dimensional $68\%$ and $95\%$~CL allowed contours from Planck temperature, polarization and lensing observations for the most relevant cosmological parameters for the $\Lambda$CDM model and for a model with 17 redshift nodes each of them associated to a different $A_{\rm \dot{\Phi}}$.}\label{fig:f18}
\end{figure*}

To demonstrate the stability of our findings, we implement a tomographic redshift analysis for $\dot{\Phi}(z)$ using the same redshift nodes as $S_{\rm ISW}(z)$. Based on the constraining value $A_{500}=1.57^{+0.44}_{-0.32}$, we still find a $\sim2\sigma$ deviation from $\Lambda$CDM in a 12-bin gravitational potential model around $z=500$. In the range $z\in[200,\,800]$, the amplitudes $A_{300}=0.58^{+0.24}_{-0.47}$ and $A_{700}=0.69^{+0.31}_{-0.46}$ also give the same evolution behavior as that in the ISW amplitude scenario. This illustrates that this $\sim2\sigma$ excess relative to $\Lambda$CDM during the dark ages era is stable (see Fig.~\ref{fig:f17}). Interestingly, the $1\sigma$ constraint $A_{10}=0.56^{+0.21}_{-0.52}$ presents a $\sim2\sigma$ deviation from $\Lambda$CDM in the low-$z$ range $[0,\,20]$. This is a different property from the case of the ISW source function. 
Moreover, we also carry out a similar tomographic redshift analysis for $\Phi(z)$ in a 17-bin model (see Fig.~\ref{fig:f18} and Tab.~\ref{tab:smt2}). The corresponding redshfit nodes are $z=0.5$, 1.5, 4,8, 30, 100, 200, 300, 400, 500, 600, 700, 800, 900, 1000, 1100 and 1200 in the range $z\in[0,\,1250]$. We obtain the constraints $A_{\Phi600}=1.53^{+0.64}_{-0.37}$ and $A_{\Phi700}=0.52^{+0.15}_{-0.51}$, which also presents a $\sim2\sigma$ deviation from the $\Lambda$CDM case in the range $z\in[200,\,800]$. This result may imply that the $2\sigma$ deviation from the standard expectation in the ISW effect amplitude is solid and originated from the anomalous behavior of the gravitational potential during the evolution of the universe.

\subsection{Cosmological tensions and ISW effect}
\begin{figure*}
	\centering
        \includegraphics[scale=0.35]{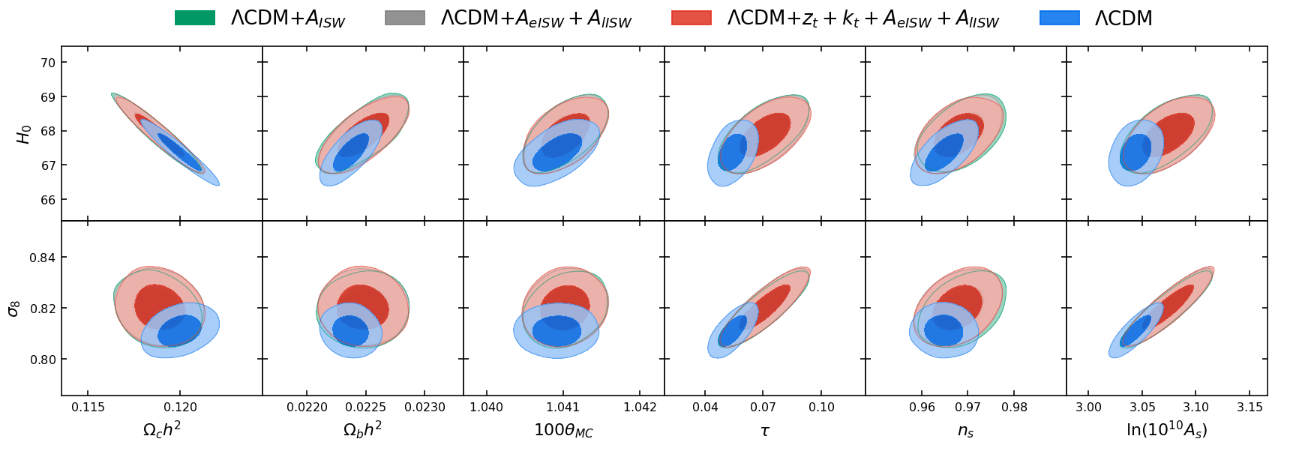}
	\caption{Two-dimensional joint posterior distributions of ($H_0$, $\sigma_8$) and six basic parameters depicting the $\Lambda$CDM from Planck temperature, polarization and lensing observations for three ISW amplitude scenarios. The $\Lambda$CDM model serves as a comparison. }\label{fig:f19}
\end{figure*}

\begin{figure*}
	\centering
	\includegraphics[scale=0.4]{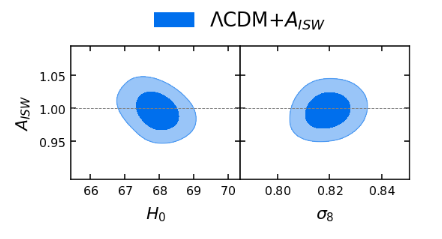}
        \includegraphics[scale=0.4]{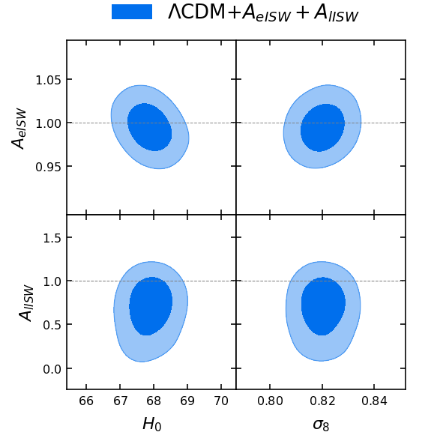}
        \includegraphics[scale=0.4]{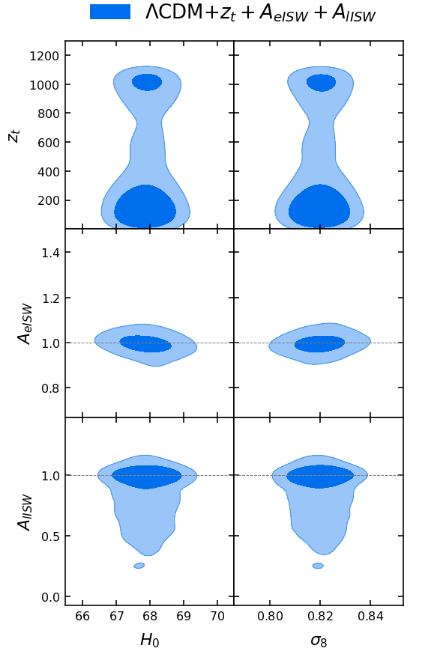}
        \includegraphics[scale=0.4]{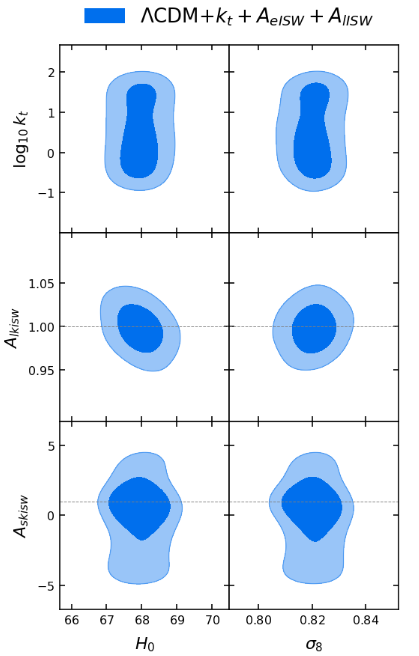}
        \includegraphics[scale=0.4]{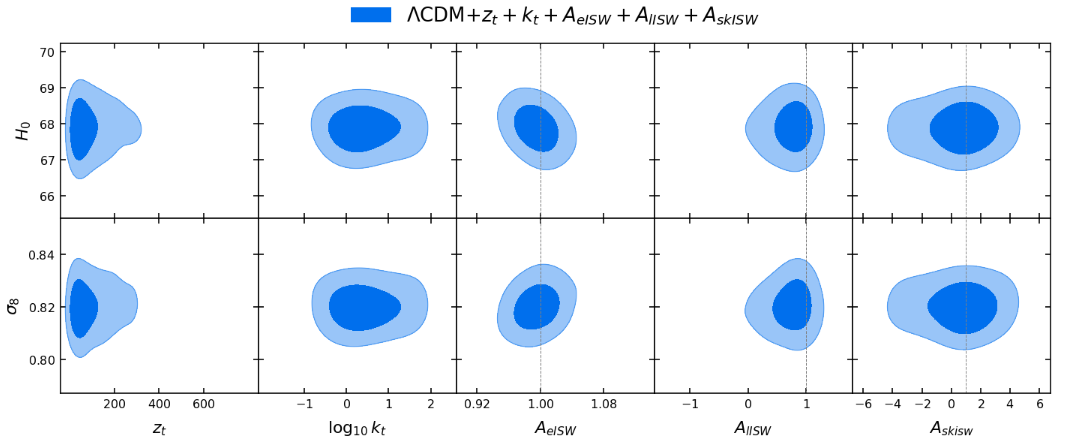}
	\caption{Two-dimensional joint posterior distributions of ($H_0$, $\sigma_8$) and ISW effect related parameters from Planck temperature, polarization and lensing observations for five ISW amplitude scenarios.}\label{fig:f20}
\end{figure*}

\begin{figure*}
	\centering
        \includegraphics[scale=0.55]{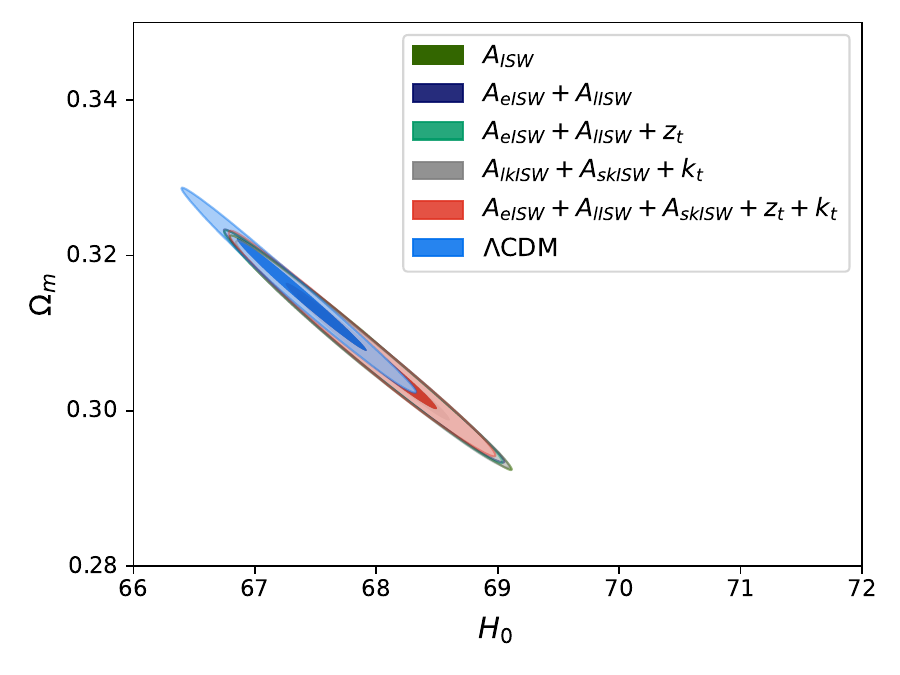}
        \includegraphics[scale=0.55]{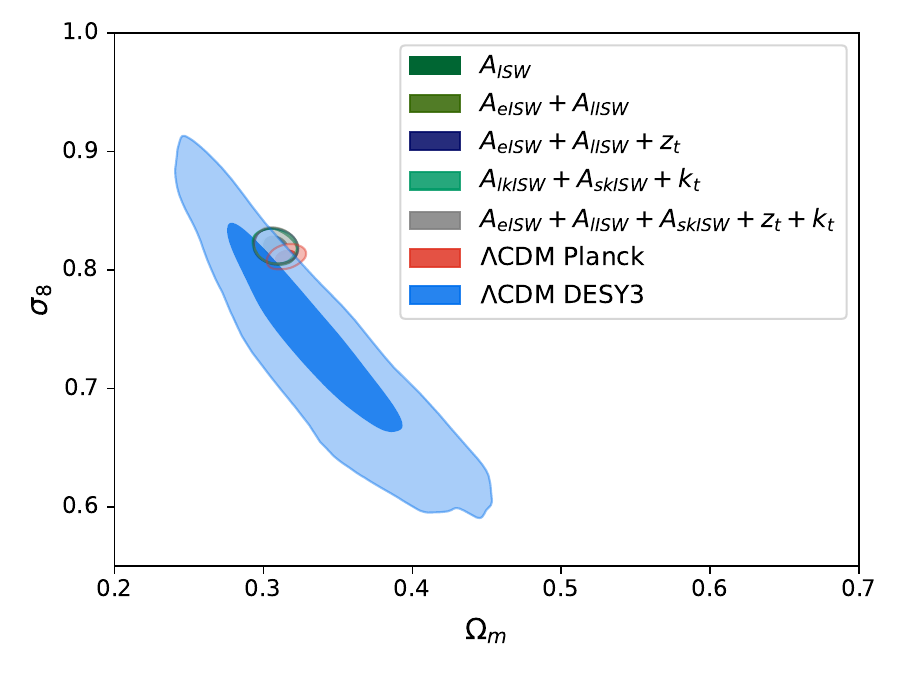}
	\caption{{\it Left.} Two-dimensional posterior distributions of the parameter pair ($H_0$, $\sigma_8$) from Planck temperature, polarization and lensing observations for five ISW amplitude scenarios and the $\Lambda$CDM model. {\it Right.} Two-dimensional posterior distributions of the parameter pair ($\Omega_m$, $\sigma_8$) from Planck temperature, polarization and lensing observations for five ISW amplitude scenarios and the $\Lambda$CDM model. We include the Dark Energy Survey Year 3 (DESY3) contour as a comparison.}\label{fig:f21}
\end{figure*}

ISW is a powerful probe of dark energy, since its source function is a combined quantity of the gravitational potential $\Phi(z)$ and of the optical depth $\tau$. The former can constrain dark energy at the perturbation level, while the latter limits its background evolution. The cosmological tensions emerging in recent years are closely linked with dark energy, especially the so-called Hubble constant ($H_0$) and matter clustering fluctuation ($\sigma_8$) ones \cite{Abdalla:2022yfr,DiValentino:2020zio}. Hence, the ISW effect may have a close connection with both tensions. As a consequence, it is interesting to study these two tensions in light of the ISW effect amplitude models we consider. The marginalized posterior distributions are shown in Figs.~\ref{fig:f19}-\ref{fig:f21}. Taking four ISW amplitude models into account, we find that they all give the same constraints on six basic parameters in $\Lambda$CDM and prefer a slightly larger $H_0$ and $\sigma_8$. It is well known that a larger spectral index $n_s$ generally leads to a larger expansion rate and a stronger matter density fluctuation. Since Planck CMB data supports a larger $n_s$ (see Fig.~\ref{fig:f19}) for all the ISW amplitude scenarios, it is straightforward to explain our findings. Furthermore, for completeness, we display the joint posterior distributions of the ISW amplitude related parameters and $H_0$ and $\sigma_8$ in five scenarios (see Fig.~\ref{fig:f20}). Unfortunately, we do not find any obvious degeneracy. In order to study both tensions more clearly, we present the Planck constraints on the parameter pairs ($H_0$, $\Omega_m$) and ($\Omega_m$, $\sigma_8$) in Fig.~\ref{fig:f21}. As expected, all the ISW amplitude models give a slightly smaller matter fraction than $\Lambda$CDM, since they prefer a slightly larger $H_0$. It is noteworthy that all the ISW amplitude models can not resolve the $\sigma_8$ tension between Planck CMB and DESY3 large scale structure data \cite{DES:2022ccp}, although they enlarge statistically the model parameter space by introducing more freedoms in the CMB temperature scalar sources.

\end{document}